\PassOptionsToPackage{dvipsnames}{xcolor}

\documentclass[galaxies,review,accept,pdftex,moreauthors]{Definitions/mdpi}

\firstpage{1} 
\pubvolume{1}
\issuenum{1}
\articlenumber{0}
\pubyear{2026}
\copyrightyear{2026}
\externaleditor{Firstname Lastname} 
\datereceived{16 July 2026} 
\daterevised{15 August 2026} 
\dateaccepted{18 August 2026} 
\datepublished{ } 

\usepackage{booktabs}
\usepackage{multirow}

\let\plainuparrow\uparrow
\renewcommand{\uparrow}{\mathrel{\textcolor{ForestGreen}{\plainuparrow}}}
\let\plaindownarrow\downarrow
\renewcommand{\downarrow}{\mathrel{\textcolor{red}{\plaindownarrow}}}
\newcommand{\violetsim}{\mathrel{\textcolor{violet}{\sim}}}

\newcommand{\msun}{M_\odot}
\newcommand{\re}{R_e}
\newcommand{\mstar}{M_\ast}

\newcommand{\sfgs}{star-forming galaxies}

\newcommand{\qgs}{quiescent galaxies}
\newcommand{\smr}{size--mass relation}
\newcommand{\smrs}{size--mass relations}

\graphicspath{{./}{cartoons_v2/}}

\Title{Size--Mass Relation Shows Its Colours: Contrasting Physical Imprints of Galaxy Evolution in Rest-Frame  
UV and Optical}

\Author{Angelo George $^{1,2,}$*\orcidA{}, 
        Marcin Sawicki $^{2}$\orcidB{}
        and Ivana Damjanov $^{2}$\orcidC{}}

\AuthorNames{Angelo George, Marcin Sawicki and Ivana Damjanov}

\address{%
$^{1}$ \quad Institute of Astronomy and Astrophysics, Academia Sinica (ASIAA), No.~1, Sec.~4, Roosevelt Rd., Taipei~106319, Taiwan\\
$^{2}$ \quad Institute for Computational Astrophysics and Department of Astronomy \& Physics, Saint Mary's University, 923~Robie Street, Halifax, NS B3H~3C3, Canada; marcin.sawicki@smu.ca (M.S.); ivana.damjanov@smu.ca (I.D.) 
}

\corres{Correspondence: angelogeorget@gmail.com
}

\abstract{The galaxy size--mass relation (SMR) is a key scaling relation used to constrain the physical processes that build galaxy structure, yet it is almost always measured in a single rest-frame optical band, where the light traces the bulk of the old stellar mass. Tracing younger populations with flux-weighted ages of $\sim$100--500 Myr and low-metallicity stars, the rest-frame near-ultraviolet opens a new stellar window on this scaling relation. Because each process redistributes the light of young and old stars differently, the same mechanism shifts the slope and zero point of the SMR by different amounts in the two wavelength regimes. Here we {review and} synthesize the effects of main physical processes on the form of the SMR for star-forming and quiescent galaxies in the rest-UV and optical. For each process, we start from its underlying physics, the galaxy stellar masses it affects, and the light it adds/removes/rearranges, anchoring the predictions to observations and simulations. We validate the predicted imprints with forward Monte Carlo modelling. The two-wavelength view breaks several degeneracies that single-band analyses cannot, most notably between minor mergers, dry major mergers, and adiabatic expansion. These results motivate joint rest-UV and optical SMR measurements with current and upcoming wide-field imaging~surveys.}


\begin{document}

\section{Introduction}\label{sec:intro}

Galaxies follow a tight power-law correlation between effective radius $\re$ and stellar mass $\mstar$ since at least $z \sim 8$~\citep{shenSizeDistributionGalaxies2003,trujilloLuminositySizeMassSizeRelations2004,vanderwel3DHST+CANDELSEvolutionGalaxy2014,langeGalaxyMassAssembly2015,mowlaCOSMOSDASHEvolutionGalaxy2019, damjanovQuiescentGalaxySize2019,damjanovSizeSpectroscopicEvolution2023,kawinwanichakijHyperSuprimeCamSubaru2021,cutlerDiagnosingDASHCatalog2022,georgeTwoRestframeWavelength2024,georgeEffectsEnvironmentSize2025,itoSizeStellarMass2024,martoranoSizeMassRelation2024,ormerodEPOCHSVISize2024, wardEvolutionSizeMass2024,westcottEPOCHSXIStructure2025}. This size--mass relation (SMR) is commonly parametrized as
\begin{equation}
\log \re \;=\; \log R_0 \;+\; \alpha \log \frac{\mstar}{M_0},
\label{eq:smr}
\end{equation}
where $R_0$ and $\alpha$ are fitting parameters (zero point and slope, respectively), and~$M_0$ is a normalization mass. Studies often choose the fiducial normalization mass to be \mbox{$M_0 = 5 \times 10^{10} \msun$} so that $R_0$ approximately gives the characteristic size of a Milky Way-massed galaxy~\citep{licquiaImprovedEstimatesMilky2015}. This fiducial mass also coincides with the knee of the galaxy stellar mass function~($\log (M^{\star}/\msun) \approx 10.7$; \citep{davidzonCOSMOS2015GalaxyStellar2017}), and~lies in the regime where both star-forming and quiescent populations are sufficiently well sampled to constrain $\alpha$ and $\log R_0$ jointly.

Star-forming galaxies (SFGs) and \qgs\ (QGs) populate distinct loci in the size--stellar mass plane. At~any given redshift and stellar mass, SFGs tend to have larger characteristic sizes than QGs except at very high masses ($M_*>10^{11}\msun$)~\citep{shenSizeDistributionGalaxies2003,vanderwel3DHST+CANDELSEvolutionGalaxy2014,kawinwanichakijHyperSuprimeCamSubaru2021,georgeTwoRestframeWavelength2024}. Although~the SMR slope is uniform for SFGs at $\log \mstar/\msun\gtrsim9$, the~slope for QGs changes at a pivot stellar mass, $\log \mstar/\msun\sim10.4$, where the QGs exhibit a steeper relation at higher masses than lower masses~\citep{mowlaMassdependentSlopeGalaxy2019, kawinwanichakijHyperSuprimeCamSubaru2021,georgeTwoRestframeWavelength2024}. The~differences in the SMR parameters $(\log R_0,\alpha)$ between SFGs, low-mass QGs and high-mass QGs arise because these parameters encode the cumulative imprint of the processes that build galaxy structure---in situ star formation, bulge growth, mergers, accretion, environmental harassment, ram-pressure and tidal stripping, and~the population-level effect of progenitor bias~(for reviews, see~ \citep{conseliceEvolutionGalaxyStructure2014,pillepichSimulatingGalaxyFormation2018,manStarFormationQuenching2018}). Tracing the evolution of the SMRs over cosmic time is therefore a useful tool for exploring the physical processes that shape~galaxies.

One of the best ways to trace galaxy evolution in the size--mass plane would be to use the mass-weighted effective radius (half-mass radius, $R_m$; \citep {moslehConnectionStellarMass2017,suessHalfmassRadii70002019}). However, measuring $R_m$ is observationally more demanding than measuring the half-light radius, $R_e$, because~the former requires the mass-to-light ratio ($M/L$) as a function of radius. Furthermore, pinning down that spatially varying $M/L$ needs spatially resolved photometry over a broad wavelength range. Measuring $R_e$ from a single-band sky image, by~contrast, is far more straightforward. A~vast majority of observational works therefore use light-weighted $R_e$ in analysing the~SMR.

However, light-weighted sizes are sensitive to the wavelength at which they are measured because~different types of stars dominate the flux density in different wavelength regimes. In~addition, the~mapping from rest-frame to the observed band is redshift-dependent. Hence, it is important to use multi-band data to trace the SMR evolution in the same rest-frame wavelength across cosmic time. 
In practice, at~least since the cosmic noon, observational works commonly infer the SMR in a single rest-frame optical/near-IR band, where light traces the old, mass-dominant stellar population~\citep{bruzualStellarPopulationSynthesis2003,conroyModelingPanchromaticSpectral2013}. 

The rest-frame near-UV is, however, a~complementary tracer: it is sensitive to a flux-weighted age of $\sim$300 Myr~\citep{calzettiLocalStarburstsPerspectives2005,bruzualStellarPopulationSynthesis2003,martinGalaxyEvolutionExplorer2005} and to low-metallicity stars~\citep{wortheyComprehensiveStellarPopulation1994,meulenaerDerivingPhysicalParameters2014}, and~thus probes a different stellar component of the same galaxies. Galaxies generally have negative colour gradients with bluer outskirts~\citep{labarberaSPIDERSampleGalaxy2010,suessHalfmassRadii70002019,moslehConnectionStellarMass2017} and UV sizes are systematically larger than optical sizes in both SFGs and QGs~\citep{georgeTwoRestframeWavelength2024,georgeEffectsEnvironmentSize2025}. Because~each physical process in galaxy evolution transforms the SMR in its own way, the~UV and optical SMRs respond to each process with varying sensitivity and, in~some cases, in~opposite directions. The~two-wavelength comparison (UV vs.\ optical) therefore breaks several degeneracies that single-band analyses are subject~to.

The motivation for this synthesis is twofold. First, a~growing body of multi-band SMR and structural-evolution work has emerged in recent years. This includes spatially resolved colour-gradient studies~\citep{labarberaSPIDERSampleGalaxy2010,moslehConnectionStellarMass2017,suessHalfmassRadii70002019,suessDissectingSizeMassS1Mass2021,salimDustAttenuationCurves2018}, bulge+disk decompositions across wavelengths~\citep{moslehGalaxySizes22020,robothamProFusePhysicalMultiband2022,hashemizadehDeepExtragalacticVisible2022,kawinwanichakijStellarMassSizeRelation2025}, and~statistical UV+optical SMR measurements at $z <1$~\citep{georgeTwoRestframeWavelength2024,georgeEffectsEnvironmentSize2025}. The~theoretical interpretations of these results illustrate how the same physical processes leave qualitatively different fingerprints in the different bands. These differences call for a unified theoretical framework in which the SMRs in rest-frame UV and optical are analysed in tandem. This can disentangle process-specific signatures far more effectively than any single wavelength band alone.  

Second, recent and upcoming multi-wavelength imaging datasets are making it possible to study the evolution of the SMR at multiple rest-frame wavelengths for very large galaxy samples. Deep U-band imaging from CFHT--CLAUDS~\citep{sawickiCFHTLargeArea2019}, combined with Subaru HSC-SSP deep optical imaging~\citep{aiharaHyperSuprimeCamSSP2018,aiharaThirdDataRelease2022}, already delivers joint U+optical photometry over $\sim$18.6 square degrees. The~Ultraviolet Near-Infrared Optical Northern Survey (UNIONS;~\citealp{gwynUNIONSUltravioletNearinfrared2025}) extends this coverage to $\sim$6250 square degrees of the northern sky at shallower depth. Looking ahead, the~Vera C.~Rubin Observatory's LSST~\citep{ivezicLSSTScienceDrivers2019}, Euclid~\citep{scaramellaEuclidPreparationEuclid2022}, and~Roman~\citep{akesonWideFieldInfrared2019} will extend the joint UV+optical SMR into the low-mass regime for $10^{6}$--$10^{7}$ galaxies. The~Euclid NISP and Roman WFI near-IR channels in particular will additionally deliver rest-frame $\sim$1 $\upmu$m sizes {for these very large samples}---a regime that traces the bulk old stellar mass even more faithfully than the rest-optical and is largely insensitive to dust attenuation. Together, these datasets extend the two-wavelength comparison developed here to a UV/optical/near-IR triplet for the same~galaxies.

In this paper we {review and synthesize} how the multi-wavelength SMR enabled by such multi-wavelength surveys can be used to discriminate between the physical processes that drive galaxy evolution. Working process by process, we ask how each mechanism redistributes UV-bright and optical-bright stellar light in individual galaxies, what trajectory this implies in the size--stellar-mass plane, and~what the population-level signature is on the slope and zero point of the SMR in each wavelength~regime.

\begin{figure}[H]
\includegraphics[width=\linewidth]{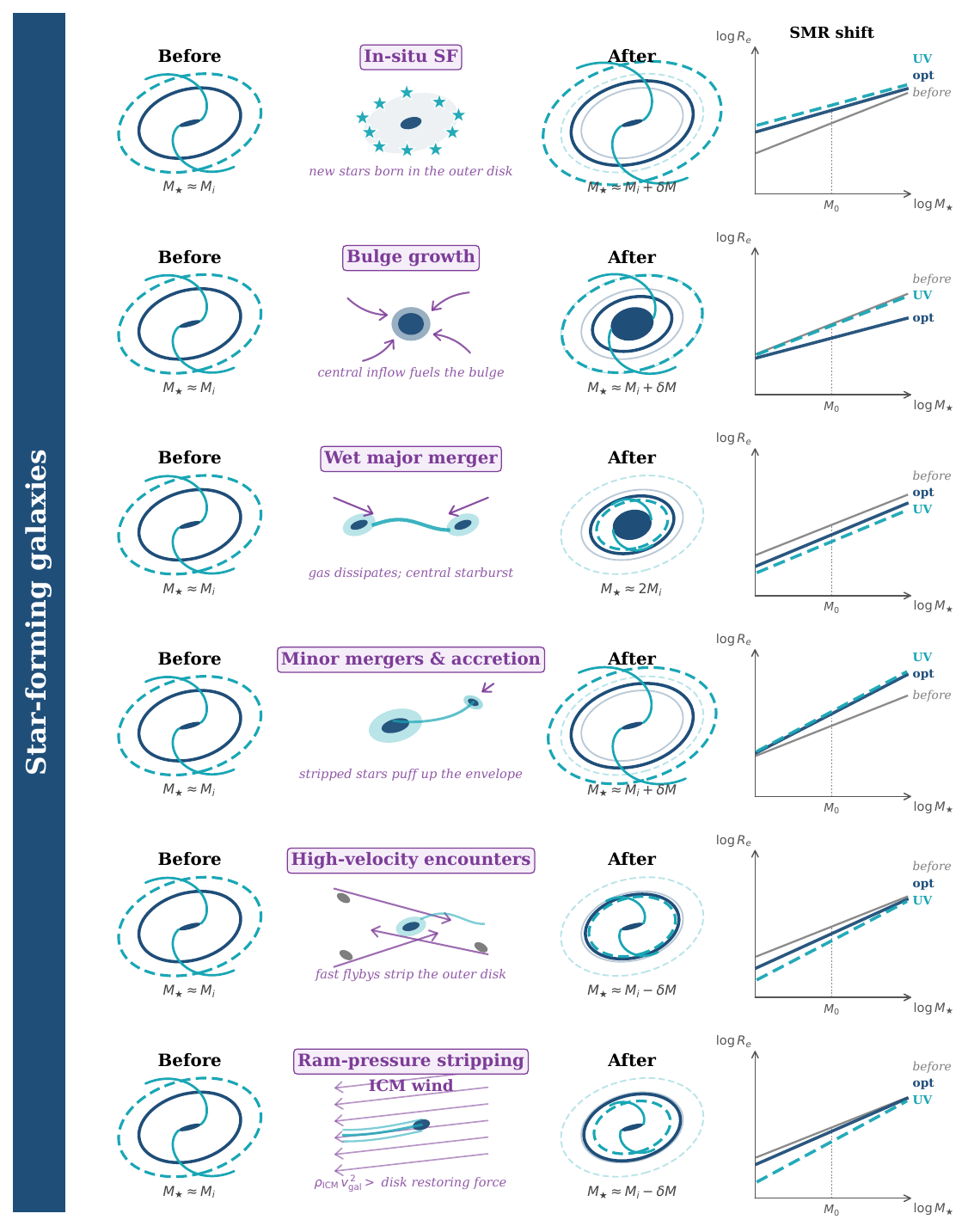}
\caption{Wavelength-dependent effects of six physical processes on the SMR of SFGs. Each row corresponds to one process and contains four panels. \emph{Before} and \emph{After} render a representative SFG as a barred-spiral icon (deep-blue bar and cyan spiral arms) enclosed by two effective-radius rings: a dashed cyan ring at $R_{\rm UV}$ and a solid deep-blue ring at $R_{\rm opt}$. The~\emph{Process} panel (header names the process) sketches the physical mechanism with purple iconography: cold-gas streams, merging galaxies, ICM wind, infalling satellites, or~radial outflows. Faint grey ghosts of the Before-state rings are overlaid on the After panel. Hence, the~size and morphology change is visible at a glance. The~rightmost panel (\emph{SMR shift}; title shown only on the top row) is a schematic of how the population \smr\ $\log\re = \log R_0 + \alpha \log(\mstar/M_0)$ moves: the grey line is the pre-process reference (shared across rows so the comparison isolates the \emph{differential} effect of each process), and~the dashed cyan and solid deep-blue lines are the post-process \smrs\ in rest-frame UV and rest-frame optical, respectively; right-anchored labels identify the lines. Signs and the relative magnitudes of the UV vs.\ optical shifts follow the literature synthesis of Section~\ref{ssec:sfgs} and Table~\ref{tab:summary}; line slopes and intercepts in the inset are illustrative, not numerical~predictions.\label{fig:sfg}}
\end{figure}

\begin{figure}[H]

\includegraphics[width=\linewidth]{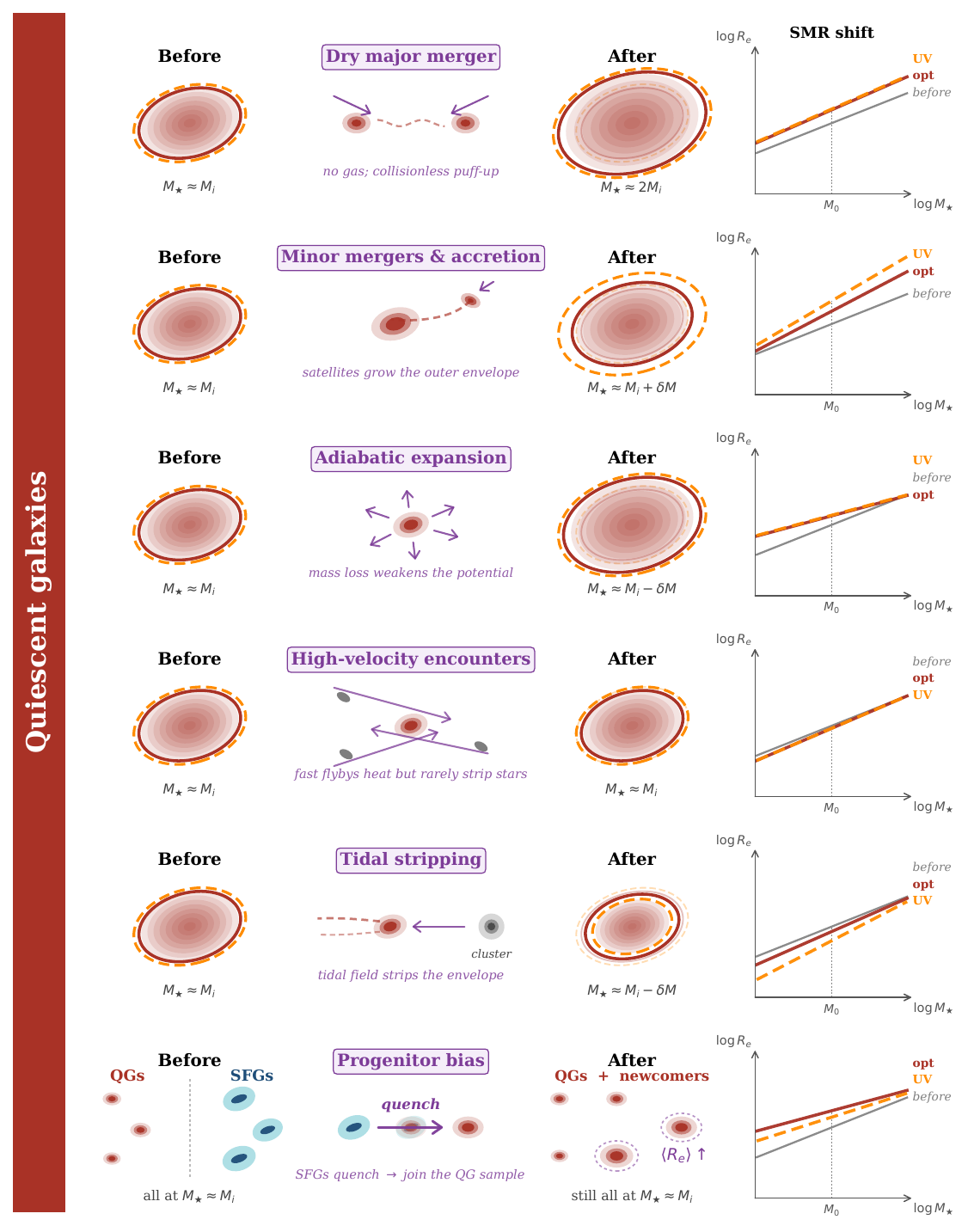}
\caption{As Figure~\ref{fig:sfg} but for QGs. The~QG host is rendered as a smooth red gradient ellipsoid; the size rings are dashed darkorange for $R_{\rm UV}$ and solid red for $R_{\rm opt}$. Process iconography includes merging galaxies, accreted satellites, post-feedback adiabatic outflows, and~tidal perturbers; the bottom row (\textit{progenitor bias}) is rendered differently because the effect is population-level rather than individual-galaxy: the \emph{Before} panel shows two groups of galaxies at the same $\mstar$ (small QGs at left, larger SFGs at right) and the \emph{After} panel shows the enlarged QG sample after the SFG cohort quenches and joins it. Hence, the~mean $\re$ at fixed $\mstar$ rises without any individual galaxy growing. The~SMR-shift convention is identical to Figure~\ref{fig:sfg} with the QG-family colours substituted (dashed orange = UV, solid red = optical). Signs and wavelength contrasts follow Section~\ref{ssec:qgs} and the lower half of Table~\ref{tab:summary}.\label{fig:qg}}
\end{figure}

This paper is organized as follows. Section~\ref{sec:framework} defines the wavelength contrast and our standard SMR baseline. Section~\ref{sec:processes} provides a process-by-process discussion, organized into a subsection for SFGs (Section~\ref{ssec:sfgs}) and a subsection for QGs (Section~\ref{ssec:qgs}). For~every process we describe (1) the process itself, (2) how it redistributes UV-bright and optical-bright stellar light in individual galaxies, (3) the trajectory of an individual galaxy in the size--mass plane, (4) the mass range in which the process is most efficient, and~(5) the resulting changes in the slope and zero point of the SMR in UV versus optical. Section~\ref{sec:caveats} discusses degeneracies (age--metallicity, dust, the~UV upturn, and~single-S\'{e}rsic versus bulge+disk biases) that complicate the empirical separation of these effects. Table~\ref{tab:summary} compiles the qualitative directions for both populations and Figures~\ref{fig:sfg} and \ref{fig:qg} visualize them. We close in Section~\ref{sec:summary} with a set of testable diagnostics for ongoing and upcoming~surveys. 

Throughout the text, we use ``UV''/``near-UV'' to mean the rest-frame near-UV at $\lambda_{\rm rest}~\approx~3000$ \AA\ and ``optical'' to mean rest-frame optical at $\lambda_{\rm rest}~\approx~5000$ \AA. The~two choices are motivated by stellar-population physics. The~$\sim$3000 \AA\ is short enough to be dominated by main-sequence A and late-B stars (flux-weighted age $\sim$$10^{8.5}$ yr) yet long enough to lie above the ``UV upturn'' contribution from extreme-horizontal-branch and post-AGB stars in old populations at $\lambda~\lesssim~2500$ \AA\ (Section~\ref{sec:caveats}). The~$\sim$5000 \AA\ samples the FGK main-sequence and the giant-branch stars that hold most of the stellar mass, and it is also the wavelength range at which the great majority of published SMRs are measured, enabling direct comparability with the~literature. 

We note that the rest-frame near-IR ($\sim$1--1.6 $\upmu$m) is {a much better tracer} of total stellar mass as~it is weighted toward K and M dwarfs and giants and much less affected by dust. Therefore, the~predictions we develop here for the optical regime apply \emph{a fortiori} to the near-IR, with~IR--UV wavelength contrasts that should be comparable to or larger than those of the optical--UV. 
Similarly, our interpretations of the rest-frame UV can be easily extended to other star formation tracers, such as the H$\alpha$ emission~line.

\begin{table}[H]
\caption{Predicted directions of the changes in the slope $\alpha$ and zero point $\log R_0$ of the galaxy \smr\ in rest-frame UV and rest-frame optical, process by process, separately for \sfgs\ and \qgs. The~signs and wavelength contrasts derive from the literature synthesized in Section~\ref{sec:processes}. Symbols: $\uparrow$ increase, $\downarrow$ decrease, $\violetsim$ negligible. ``UV $>$ Opt'' denotes that the \emph{magnitude} of the shift is larger in UV than in optical; sign columns indicate the direction in each wavelength separately. The~last column grades the population-averaged amplitude of each process over $z=1 \to 0$ from~order-of-magnitude estimates, which combine the measured rates, timescales, affected population fractions, and~size--mass-plane tracks compiled in Section~\ref{sec:processes}.\label{tab:summary}}
\begin{adjustwidth}{-\extralength}{0cm}
\begin{tabularx}{\fulllength}{llccccccl}
\toprule
\textbf{Pop.} & \textbf{Process} & \multicolumn{2}{c}{$\boldsymbol{\Delta\alpha}$} & & \multicolumn{2}{c}{$\boldsymbol{\Delta\log R_0}$} & \textbf{Wavelength Contrast} & \textbf{Amplitude} \boldmath{$^{a}$} \\
\cmidrule{3-7}
 & & \textbf{UV} & \textbf{Opt} & & \textbf{UV} & \textbf{Opt} & & \\
\midrule
\multirow{6}{*}{SFG}
 & In situ SF/inside-out disk growth & $\downarrow$ & $\downarrow$ & & $\uparrow$ & $\uparrow$ & UV $\gtrsim$ Opt & significant \\
 & Bulge growth/inside-out quenching & $\violetsim$ & $\downarrow$ & & $\violetsim$ & $\downarrow$ & UV $<$ Opt & modest \\
 & Wet major mergers                   & $\violetsim$ & $\violetsim$ & & $\downarrow$ & $\downarrow$ & UV $>$ Opt $^{d}$ & modest \\
 & Minor mergers \& accretion          & $\uparrow$ & $\uparrow$ & & $\uparrow$ & $\uparrow$ & UV $\sim$ Opt & modest \\
 & High-velocity encounters                   & $\uparrow$ & $\uparrow$ & & $\downarrow$ & $\downarrow$ & UV $>$ Opt & negligible $^{b}$ \\
 & Ram-pressure/tidal stripping      & $\uparrow$ & $\uparrow/\violetsim$ & & $\downarrow$ & $\downarrow/\violetsim$ & UV $>$ Opt & negligible $^{b}$ \\
\midrule
\multirow{6}{*}{QG}
 & Dry major mergers                   & $\uparrow$ & $\uparrow$ & & $\uparrow$ & $\uparrow$ & UV $\sim$ Opt & modest \\
 & Minor mergers \& accretion          & $\uparrow$ & $\uparrow$ & & $\uparrow$ & $\uparrow$ & UV $>$ Opt & significant \\
 & Adiabatic expansion                 & $\downarrow$ & $\downarrow$ & & $\uparrow$ & $\uparrow$ & UV $\sim$ Opt & negligible $^{c}$ \\
 & High-velocity encounters            & $\violetsim$ & $\violetsim$ & & $\violetsim$ & $\violetsim$ & UV $\sim$ Opt & negligible $^{b}$ \\
 & Tidal stripping                     & $\uparrow$ & $\uparrow/\violetsim$ & & $\downarrow$ & $\downarrow/\violetsim$ & UV $>$ Opt & negligible $^{b}$ \\
 & Progenitor bias                     & $\downarrow$ & $\downarrow$ & & $\uparrow$ & $\uparrow$ & UV $<$ Opt & significant \\
\bottomrule
\end{tabularx}
\end{adjustwidth}
\noindent\footnotesize{Notes: The predictions are for galaxies above the QG SMR pivot mass ($\log\mstar/\msun \gtrsim 10.4$) and SFGs above $\log\mstar/\msun \gtrsim 9.5$. Sign symbols apply qualitatively to the steady-state effect of each process averaged over the population on which it is most efficient. Because~mergers and stripping events are stochastic, all of those entries also \emph{broaden} the intrinsic scatter of the SMR, in~addition to shifting its slope and intercept. Throughout this table, $\Delta\log R_0$ denotes the shift in mean $\log\re$ at the population fiducial mass ($5\times10^{10} \msun$), not the formal intercept of the linear fit extrapolated to $\log\mstar = 0$; this convention keeps the sign of $\Delta\log R_0$ aligned with the direction of size change at the masses where each process~operates. $^{a}$ ``Significant'': comparable to or exceeding typical SMR measurement uncertainties ($|\Delta\alpha| \gtrsim 0.05$ or $|\Delta\log R_0| \gtrsim 0.04$ dex over $z=1 \to 0$); ``modest'': within a factor of a few below them; ``negligible'': an order of magnitude or more below them. Grades additionally discount amplitudes whose sign is not robust to parameter choices, transient or upper-bound estimates, and~partial cancellation between co-acting~channels. $^{b}$ Population-diluted: the per-galaxy effect is substantial, but~the effect is negligible over the whole population as only galaxies in dense environments are affected; the predicted signatures are recoverable in cluster and dense-group~samples. $^{c}$ Confined to the quenching epoch: shapes the arrival sizes of newly quenched galaxies (mainly at $z \gtrsim 1.5$) rather than driving continued evolution of the established quiescent~population. $^{d}$ Transient: applies during the $\lesssim$1 Gyr starburst phase; the time-averaged population contrast is small, and~its $z=1 \to 0$ change is mildly positive (the UV zero point recovers as the merger rate~declines).}
\end{table}

\section{Framework: The Two-Wavelength~SMR}\label{sec:framework}

Rest-frame UV light is dominated by main-sequence A and late-B stars and by the outer envelopes of intermediate-mass stars~\citep{bruzualStellarPopulationSynthesis2003,calzettiLocalStarburstsPerspectives2005,conroyModelingPanchromaticSpectral2013}, while the rest-frame optical is dominated by the ensemble of FGK main-sequence and giant-branch stars that hold most of the stellar mass. For~a single stellar population this difference translates into a flux-weighted age of $\sim$$10^{8.5}$ yr in the UV versus $\sim$$10^{9.5}$ yr in the optical~\citep{bruzualStellarPopulationSynthesis2003,conroyModelingPanchromaticSpectral2013}. For~a composite population with ongoing or recent star formation, the~UV light is preferentially produced in regions of high specific star-formation rate (sSFR), while the optical light tracks the spatial distribution of the bulk of the old stellar~mass.

Two consequences follow from this wavelength contrast. First, any process that {changes where the most intense star formation occurs} (such as in situ growth, inside-out quenching, harassment-induced starbursts, wet mergers;~\citep{nelsonWhereStarsForm2016,tacchellaEvidenceMatureBulges2015,mooreGalaxyHarassmentEvolution1996,barnesTransformationsGalaxiesII1996}) will reshape the UV light profile---and therefore the UV-derived $\re$---more than the optical profile. Second, any process that adds, removes, or~kinematically rearranges the bulk old stellar mass, which alters its radial population mix (dry major mergers, adiabatic expansion, deep tidal stripping of the relaxed component;~\citep{naabMinorMergersSize2009,fanDramaticSizeEvolution2008,merrittRelaxationTidalStripping1983}), will reshape the optical and UV profiles by comparable amounts. One further channel modulates this dichotomy: at fixed old age, metal-poor populations remain UV-brighter~\citep{wortheyComprehensiveStellarPopulation1994,meulenaerDerivingPhysicalParameters2014}. Any process that builds up or strips the metal-poor, ex situ outer envelopes of QGs (minor mergers/accretion and tidal stripping; Sections~\ref{sssec:minorQG} and~\ref{sssec:tidalQG}) carry a UV-side bias even in the absence of young stars. Even before any process acts, a~typical galaxy's negative colour gradient makes its UV size exceed its optical size (we return to this topic in Section~\ref{sec:caveats}. The~differential response of the UV and optical SMRs to a given physical process is what we denote here as its ``wavelength contrast''. In~Section~\ref{sec:processes} we work through twelve such processes; their predicted wavelength contrasts are compiled in Table~\ref{tab:summary} and visualized in Figures~\ref{fig:sfg} (for SFGs) and~\ref{fig:qg} (for QGs).

To make this differential response transparent, in~each panel of Figures~\ref{fig:sfg} and \ref{fig:qg} a grey reference line shows an idealized SMR before the physical process acts on the population; the post-process SMRs in rest-UV (dashed) and rest-optical (solid) are overlaid in the colours of the population (cyan/blue for SFGs, orange/red for QGs). In~reality, the~UV and optical SMRs of the unmodified population would themselves differ at the $\lesssim$0.1 dex level, but~using a single shared baseline across panels isolates \emph{the differential effect of the processes on the SMR in two wavelengths}, which is the quantity of~interest.

In this synthesis, we focus on the well-measured, well-sampled part of each SMR, namely $\log\mstar/\msun~\gtrsim~9.5$ for SFGs and $\log\mstar/\msun~\gtrsim~10.4$ (the pivot mass) for QGs. We reach below these limits only where a process acts specifically there, as~with the dwarf and sub-$M^\star$ systems shaped by environmental~stripping.

\section{Process-by-Process~Synthesis}\label{sec:processes}

In this section we work through twelve physical processes that shape galaxy structure---six for \sfgs\ in Section~\ref{ssec:sfgs} (illustrated in Figure~\ref{fig:sfg}) and six for \qgs\ in Section~\ref{ssec:qgs} (illustrated in Figure~\ref{fig:qg})---and ask, for~each, how its effect on the \smr\ differs between rest-frame UV and rest-frame optical. Within~each population the six processes span three physical categories: in situ growth and quenching, mergers and accretion, and~environmental transformation. The~qualitative wavelength contrasts for all twelve processes are compiled in Table~\ref{tab:summary}.

To establish which of these processes matter at the population level, we accompany each with a qualitative amplitude estimate over a common $z=1 \to 0$ baseline, which is obtained by combining its measured occurrence rate, timescale, affected population fraction, and~size--mass-plane track from the literature, and~cross-checked with simple Monte Carlo population models (see Appendix~\ref{appsec:models} for details). The~resulting amplitude grades appear in the last column of Table~\ref{tab:summary}.

A recurring result of these estimates is that several processes carry robust per-galaxy signatures (high-velocity encounters, ram-pressure and tidal stripping) yet appear strongly diluted in a fit to the global galaxy population, because~only a small fraction of galaxies inhabit the dense clusters and groups where they operate. This dilution is an artefact of mixing environments in a single fit rather than a sign that the processes are weak. Once the sample is split by environment, these same processes become considerable in the dense-environment satellite subpopulation. 
The channels that act on the general population (in situ growth, mergers and accretion, progenitor bias), by~contrast, dominate the evolution of the fitted relations for the full~sample.


\subsection{Star-Forming~Galaxies}\label{ssec:sfgs}

\subsubsection{In-Situ Star Formation and Inside-Out Disk~Growth}\label{sssec:insituSFG}

Cosmologically averaged star formation in disk-dominated SFGs proceeds inside-out: the radial profile of star formation rate (SFR) surface density is shallower than that of stellar-mass surface density (equivalently, the~sSFR rises outward). Star formation deposits new stars at progressively larger radii with cosmic time. This is supported observationally by spatially resolved $H\alpha$ maps from 3D-HST~\citep{nelsonSpatiallyResolvedHa2012,nelsonWhereStarsForm2016,nelsonSpatiallyResolvedDust2016}, by~stacked CANDELS profiles~\citep{vandokkumAssemblyMilkyWayGalaxies252013,vandokkumFormingCompactMassive2015,wuytsCANDELS3DHSTSynergyResolved2013}, by~IFS surveys such as MaNGA and SAMI~\citep{ibarra-medelSDSSIVMaNGA2016,garcia-benitoSpatiallyResolvedStar2017,ellisonStarFormationBoosted2018}, and~by colour-gradient analyses~\mbox{\citep{suessHalfmassRadii70002019,moslehConnectionStellarMass2017}}. Analytic ``equilibrium'' models also predict the same inside-out behaviour. In~these models the SFR is set by the balance between cosmological gas inflow\footnote{This late-time, high-angular-momentum supply is delivered largely by cold accretion from the cosmic web~\mbox{\citep{keresHowGalaxiesGet2005,sancisiColdGasAccretion2008,dekelColdStreamsEarly2009}}, whose most extreme structural signature is the extended-UV (XUV) disk, where GALEX reveals young stars forming well beyond the optical edge of nearby SFGs~\citep{pazDiscoveryExtendedUltraviolet2005,thilkerSearchExtendedUltraviolet2007}}, star formation, and~outflows. Gas accreted at later times carries higher specific angular momentum and therefore settles and~forms stars at~progressively larger radii, while the central regions are assembled early from the low-angular-momentum gas accreted first, supplemented by gas transported inward through the disk~\citep{lillyGasRegulationGalaxies2013,pezzulliAccretionRadialFlows2016}. 

Because star formation proceeds inside-out, new young stars appear preferentially in the outer disk, where they radiate strongly in the rest-UV. This freshly added light broadens the UV emission, and~the UV $\re$ grows rapidly. The~rest-optical traces the older and more centrally concentrated disk, and~its $\re$ grows more slowly, lagging modestly behind the UV. {However, secular radial migration works in the opposite direction to inside-out growth, carrying old, optically bright but UV-faint stars outward through repeated scattering by spiral arms~\citep{radburn-smithOuterdiskPopulationsNGC2012,yoachimSpatiallyResolvedSpectroscopic2012,sellwoodSecularEvolutionDisk2014}. This builds an extended outer disk that grows the optical size slightly and narrows the usual excess of UV size over optical.}

In the size--mass plane an individual SFG therefore moves up and to the right, with~a steeper trajectory if the sizes are measured based on rest-frame UV rather than optical light profiles. Because~the main-sequence sSFR decreases with stellar mass~\citep{speagleHighlyConsistentFramework2014,whitakerConstrainingLowmassSlope2014,schreiberHerschelViewDominant2015,meridaProbingStarFormation2026}, the~fractional size growth is larger for SFGs with lower masses. Thus, the~low-mass end of the relation is lifted more than the high-mass end and the slope of the SMR flattens. The~flattening is, however, of~comparable magnitude in the two bands: the offset between the growth track and the relation is nearly band-independent, because~the ratio of the recent star formation extent to the stellar extent itself increases with stellar mass~\citep{nelsonSpatiallyResolvedDust2016,vanderwel3DHST+CANDELSEvolutionGalaxy2014}, compensating for the heavier UV weighting of the newly formed stars. Net effect: both slopes flatten by comparable, modest amounts; the UV zero point rises at least as fast as the optical zero point. Thus, the~baseline UV--optical characteristic size offset of the population ($\sim$0.1 dex) is maintained or grows slowly over~time.

\subsubsection{Bulge Growth and Inside-Out~Quenching}\label{sssec:bulgegrowth}

In SFGs, bulges are built initially by star formation. Gas driven inward by secular bar and spiral-arm torques~\citep{athanassoulaNatureBulgesGeneral2005,sellwoodSecularEvolutionDisk2014} and by gas-rich minor mergers~\citep{hopkinsDiscriminatingPhysicalProcesses2010,brooksBulgeFormationMergers2016} fuels a compact central starburst that grows a dense stellar core. At~high redshift, the~same central starburst is driven still more strongly by wet compaction of gas-rich disks~\citep{dekelColdStreamsEarly2009,dekelWetDiscContraction2014,zolotovCompactionQuenchingHighz2015,tacchellaEvolutionDensityProfiles2016}, typically near $\mstar \sim 10^{9.5} \msun$ at $z \sim 2$--4, whereas at $z < 1$ it is the secular and gas-rich merger channels that dominate. Once this core becomes dense enough, it stabilizes the residual gas disk against fragmentation and shuts off star formation from the inside out (morphological quenching;~\citep{martigMorphologicalQuenchingStar2009,saintongeImpactInteractionsBars2012,sachdevaGrowthBulgesDisk2017,hashemizadehDeepExtragalacticVisible2022}). Whereas bar and spiral torques mainly grow a disky pseudobulge, the~merger and compaction channels build the old, dispersion-supported classical bulge whose stars come to dominate the galaxy's central optical~light.

Old, metal-rich \emph{classical} bulges~\citep{kormendySecularEvolutionFormation2004,moorthyStellarPopulationsBulges2006,barbuyChemodynamicalHistoryGalactic2018} contribute disproportionately in the optical. They are dimmer in UV both because of their stellar age and because the sites of ongoing star formation are pushed outward, into~the disk~\citep{tacchellaEvidenceMatureBulges2015,belliMOSFIRESpectroscopyQuiescent2019,estrada-carpenterCLEARMorphologicalEvolution2023}. As~the bulge grows, the~global single-S\'{e}rsic optical $\re$ of the host galaxy shrinks (the central concentration increases and the outer disk is de-emphasized), while the UV $\re$ shrinks much less or, in~the inside-out-quenching limit, can briefly remain steady or grow as star formation migrates outward~\mbox{\citep{vandokkumFormingCompactMassive2015,tacchellaDustAttenuationBulge2018,suessDissectingSizeMassS1Mass2021,georgeTwoRestframeWavelength2024}}.

In the size--mass plane an individual galaxy therefore moves down and slightly to the right in the optical (modest central mass growth, shrinking global $\re$) but nearly horizontally in the UV. However, bulges are more prominent in massive SFGs than in low-mass systems~\citep{bluckBulgeMassKing2014,tacchellaMorphologyStarFormation2019}, and~the central 1 kpc stellar-mass surface density $\Sigma_1$, a~measure of central concentration, is highest in the most massive disks~\citep{whitakerPredictingQuiescenceDependence2017,estrada-carpenterCLEARMorphologicalEvolution2023,georgeTwoRestframeWavelength2024}. Bulge-driven compaction therefore acts preferentially at $\log\mstar/\msun  \gtrsim 10.5$. Net effect: the optical zero point drops while the UV zero point is nearly unchanged; the optical slope flattens measurably while the UV slope is nearly unchanged; and the UV--optical size \emph{difference} of SFGs grows because the central concentration becomes higher in the optical than in the~UV. 

\subsubsection{Wet Major~Mergers}\label{sssec:wet-mergers}

Gas-rich major mergers (mergers between galaxies of comparable masses; mass ratio above 1:4) between SFGs torque cold gas to the centre on $\lesssim$$10^9$ yr timescales. Such an event ignites a compact nuclear starburst, building a young classical bulge, and~leaving the remnant more centrally concentrated than either progenitor~\citep{barnesDynamicsInteractingGalaxies1992,barnesTransformationsGalaxiesII1996,springelFormationSpiralGalaxy2005,hopkinsCosmologicalFrameworkCoEvolution2008,hopkinsHowDisksSurvive2009,wellonsFormationMassiveCompact2015,wellonsDiverseEvolutionaryPaths2016,cibinelEarlyLatestageMergers2019}.

The nuclear starburst is intensely UV-bright but also dust-obscured~\citep{sandersLuminousInfraredGalaxies1996,lonsdaleUltraluminousInfraredGalaxies2006,calzettiCalibrationMidInfraredStar2007}, while the longer-lived optical light remains dominated by the disrupted progenitor disks. The~UV light therefore tracks the compact nuclear starburst more closely than does the extended optical envelope. Hence, the~UV $\re$ drops further than the optical $\re$~\citep{hopkinsCosmologicalFrameworkCoEvolution2008,wellonsFormationMassiveCompact2015,wuytsSizesKinematicsGradients2010,cibinelEarlyLatestageMergers2019}. However, in~the most heavily obscured phases, dust can suppress the nuclear UV and temporarily weaken this signature (Section~\ref{sec:caveats}).

In the size--mass plane the remnant moves down and to the right (stellar mass roughly doubles in a single event, $\re$ decreases). 
The galaxy major-merger rate at $z < 1$ is weakly mass-dependent above $\log\mstar/\msun \sim 10$~\citep{lotzMajorMinorGalaxy2011,xuMajorMergerGalaxyPairs2012,conseliceGalaxyFormationCosmological2014}, while the cold-gas fraction---and with it the dissipative shrinkage per event---decreases with stellar mass~\citep{hopkinsCosmologicalFrameworkCoEvolution2008,catinellaXGASSTotalCold2018,tacconiPHIBSSUnifiedScaling2018,janowieckiXGASSColdGas2020}. These two mass dependences act on the fitted slope in opposite directions (gas-rich low-mass remnants shrink more per event, while the higher event rate of massive galaxies transports compacted remnants up-mass) and largely cancel, and~hence, the~slope of the SMR is barely affected; the dominant effect is on the zero point. Two population-level effects further temper the wavelength signature. Firstly, the~UV-compact starburst phase lasts only \mbox{$\lesssim$0.5--1~Gyr}. Therefore, few SFGs exhibit it at any given epoch (and heavy nuclear obscuration can temporarily reverse it; Section~\ref{sec:caveats}). Secondly, a~fraction of remnants quenches rapidly and exits the star-forming sample altogether, transferring the most compact merger products to the quiescent relation~\citep{ellisonGalaxyMergersCan2022,quaiInterconnectionGalaxyMergers2023}. Net effect: both slopes are nearly unchanged, both zero points fall, and~the time-averaged zero-point drop is only slightly larger in UV than in~optical.

\subsubsection{Minor Mergers and~Accretion}\label{sssec:minorSFG}

A minor merger (mass ratio below 1:4)  delivers gas and stars to different places. The~gas sinks to the centre and feeds the bulge~\citep{hopkinsDiscriminatingPhysicalProcesses2010,brooksBulgeFormationMergers2016,kavirajImportanceMinormergerdrivenStar2014}. In~contrast, the~satellite's stars are collisionless and sink slowly. These accreted stars settle in the outer envelope building a diffuse stellar halo and increase the size~\citep{naabFormationEarlyTypeGalaxies2007,naabMinorMergersSize2009,oserTwoPhasesGalaxy2010,oserCosmologicalSizeVelocity2012,hilzHowMinorMergers2013,monachesiGHOSTSSurveyII2016,streichExtragalacticArcheologyGHOSTS2016,williamsGrowthGalaxyStellar2025}. Unlike a dry merger that drives only the envelope channel (Section~\ref{sssec:minorQG}), a~gas-rich merger, typical of SFGs, drives both. Dry mergers are approximately dissipationless (i.e., their mechanical energy is conserved rather than radiated away). A~combination of the virial theorem with this energy conservation gives the scaling relation for the final product: $\re \propto \mstar^{2}$~\citep{naabMinorMergersSize2009,hilzHowMinorMergers2013}. SFG disks, on~the other hand, fall short of this $\re \propto \mstar^{2}$ track because~gas is diverted to the centre and the diffuse debris adds little light-weighted $\re$ (ex situ fractions $\sim$5--15\%;~\citep{rodriguez-gomezStellarMassAssembly2016}).

In SFGs, the~wavelength contrast is modest, because~minor mergers grow the UV and optical $\re$ by comparable amounts. When a satellite is disrupted, its stars are tidally stripped and spread smoothly through the host's outer envelope, which is a~diffuse addition of light at large radii rather than discrete clumps. Light added at large radii enlarges $\re$ in~both the UV and the optical. What sets the contrast between the two bands is the colour of that added envelope light relative to the host's existing outskirt, not how bright the outskirt already is. In~a QG, the~outskirt is old and red, and~the accreted stars are relatively bluer. They add proportionally more UV light at large radii, and~the UV $\re$ grows far more than the optical (Section~\ref{ssec:qgs}). In~an SFG, by~contrast, the~outer disk is already blue and the added light {either matches it in colour or is slightly redder}. Hence, the~two bands then grow alike. The~contrast, if~any, runs opposite to the QG case, because~the infalling gas feeds a compact blue starburst that draws the UV $\re$ inward and leaves the optical $\re$ marginally the larger of the two~\citep{hopkinsDiscriminatingPhysicalProcesses2010,brooksBulgeFormationMergers2016,kavirajImportanceMinormergerdrivenStar2014}.

Minor-merger-driven growth is more efficient for high-mass hosts, both because their merger rate is higher~\citep{hopkinsMergersBulgeFormation2010,bluckStructuresTotalMinor2012,vandokkumGrowthMassiveGalaxies2010,tacchellaDustAttenuationBulge2018} and because their accreted-to-in situ ratio rises with mass (shown most clearly for massive centrals by \citep{pillepichHaloMassAssembly2014}; and extended to disk-host morphologies by \citep{rodriguez-gomezStellarMassAssembly2016}; see also \citep{pulsoniStellarHalosETGs2021,williamsGrowthGalaxyStellar2025}). Hence, this process \emph{steepens} the slope of the SFG SMR. Net effect: a small but mass-dependent increase in zero point and slope, in~both wavelengths to a similar~degree.

\subsubsection{High-Velocity~Encounters}\label{sssec:harassmentSFG}

High-velocity flyby encounters with other cluster members and the cluster potential (historically termed galaxy harassment) heat stellar disks, strip loosely bound material, and~can drive gas inwards to ignite a short-lived central starburst~\citep{mooreGalaxyHarassmentEvolution1996,mooreMorphologicalTransformationGalaxy1998,mooreSurvivalDestructionSpiral1999,mastropietroMorphologicalEvolutionDiscs2005,boselliEnvironmentalEffectsLateType2006,smithSensitivityHarassmentOrbit2015,lamarcaGalaxyPopulationsHydra2022}. Unlike a merger, the~galaxies move too fast to coalesce, and~it is the cumulative effect of many such impulsive shocks that reshapes the~galaxy.

The UV traces the loosely bound young outer disk, which the encounters strip first. That truncation makes the UV profile more compact, and~with no fresh gas to replenish it, the~smaller UV size persists. A~tidally triggered nuclear starburst can shrink the UV size further still, but~only transiently, fading within a few hundred Myr~\citep{mossHaSurveyEight2000}. The~optical light responds more weakly because~the old stars that dominate it sit in a more centrally concentrated profile and are tightly bound. Some of these old stars also populate the outer disk, and~stripping it removes optical light at a large radius, contracting the optical size modestly, much less than the UV. The~heating also thickens the survivors, but~that puffing is mostly vertical and does not shrink the projected optical size~further.

An individual SFG shifts to slightly lower stellar mass and a smaller size, with~the UV shrinking more than the optical. The~stripping that truncates the outer disk beats the heating that inflates it, and~the net size falls. The~effect is largest for low-mass, extended SFGs, which have the shallowest potentials and most loosely bound disks. The~observational evidence for this is a size offset confined to the low-mass end, where cluster SFGs are measurably smaller than their field counterparts, with~no such offset at higher mass~\citep{maltbyEnvironmentalDependenceStellarmasssize2010}. Even so, this strong shrinkage is reached only on the rare orbits that cross the dense cluster core~\citep{smithSensitivityHarassmentOrbit2015}. Net effect: within pure cluster samples, both slopes steepen and both zero points fall, with UV more than optical. Averaged across the SFG population as a whole, these shifts are~negligible.

\subsubsection{Ram-Pressure and Tidal~Stripping}\label{sssec:ramstripSFG}

Two distinct mechanisms strip infalling cluster SFGs. Ram pressure acts hydrodynamically, as~the intracluster medium sweeps the cold gas from the disk and removes the fuel for future star formation, while leaving the existing stars in place~\citep{gunnInfallMatterClusters1972,vollmerRamPressureStripping2001,corteseStrongTransformationSpiral2007,vollmerRamPressureStripping2012,boselliGALEXUltravioletVirgo2014,jaffeGASPIXJellyfish2018,robertsLoTSSJellyfishGalaxies2022}. Tidal stripping acts gravitationally, as~tides from massive neighbours and the cluster potential pull loosely bound stars and dark matter from the outer envelope~\citep{merrittRelaxationTidalStripping1983,aguilarDensityProfilesTidally1986,gnedinTidalEffectsClusters2003,blomSLUGGSSurveyNew2014,lokasTidalEvolutionGalaxies2020,mayesContributionStrippedNuclei2021,yiFormationPathwaysCompact2025}.

Because cold gas (and the young stars it would form) sits preferentially in the disk outskirts, ram-pressure stripping truncates the UV-bright periphery first, leaving the older optical disk and bulge largely intact~\citep{koopmannHaMorphologiesEnvironmental2004,corteseStrongTransformationSpiral2007,boselliGALEXUltravioletVirgo2014,jaffeGASPIXJellyfish2018,robertsLoTSSJellyfishGalaxies2022}. Tidal disruption, meanwhile, removes a modest fraction of the loosely bound outer stellar envelope along with the surrounding subhalo---for low-mass satellites on deep-pericentre orbits the \emph{total} (dark-matter-dominated) mass loss is severe ($\sim$80--90\%, with~the inner stellar cores far more resilient)~\citep{ghignaDensityProfilesSubstructure2000,kravtsovTumultuousLivesGalactic2004}. The~net result for an individual stripped SFG is a disproportionately larger reduction in UV size than in optical~size.

Stripping is most efficient for low-mass galaxies, which have a smaller restoring force, and~for extended ones~\citep{boselliOriginDwarfEllipticals2008,maltbyEnvironmentalDependenceStellarmasssize2010,brownColdGasStripping2017}. It therefore steepens the SMR slope and lowers the zero point, more strongly in the UV. The~SMR changes in the optical band share the same signs but stay weak, because~ram pressure removes gas rather than the old stars that determine the optical size. These shifts become negligible in pure gas-only stripping, and~appear only when the accompanying tidal forces also strip stars or the quenched outer disk fades. Net effect: the UV slope steepens and the UV zero point drops, while the optical follows weakly or not at all. The~wavelength contrast is large, and~the UV-side drop can persist for $\sim$$10^9$ yr, set by the main-sequence lifetimes of the A and late-B stars that dominate the rest-frame UV. Few SFGs are being stripped at any epoch, and~those fully stripped quench and leave the star-forming sample within $\sim$1 Gyr. Hence, as~with high-velocity encounters, the~effect on general population is small, though prominent, in cluster samples.

\subsection{Quiescent~Galaxies}\label{ssec:qgs}

\subsubsection{Dry Major~Mergers}\label{sssec:majorQG}

Mergers between gas-poor, spheroid-dominated QGs are the canonical pathway to growing the most massive ellipticals~\citep{naabPropertiesEarlyTypeDry2006,naabMinorMergersSize2009,bezansonRelationCompactQuiescent2009,vandokkumGrowthMassiveGalaxies2010,trujilloDissectingSizeEvolution2011,newmanCanMinorMerging2012,mancillasProbingMergerHistory2019}. For~an equal-mass dry merger, a~simple virial-theorem argument predicts that the remnant roughly doubles both its stellar mass and its half-light radius. The~galaxy therefore climbs a merger track of unit slope, $\mathrm{d}\log\re/\mathrm{d}\log\mstar \approx 1$, confirmed by controlled $N$-body experiments~\citep{naabMinorMergersSize2009,hilzHowMinorMergers2013}. 

Because both progenitors are old, the~merger does not strongly redistribute UV-bright versus optical-bright stellar light: UV and optical $\re$ grow by similar amounts. The~wavelength contrast is therefore small, except~for a residual UV-upturn contribution from extreme-horizontal-branch and post-AGB stars in the most massive QGs (Section~\ref{sec:caveats}).

During the merger, a~single galaxy climbs up and to the right along its own merger track, not along the population SMR. This track runs at $\mathrm{d}\log\re/\mathrm{d}\log\mstar \approx 1$, steeper than the relation ($\alpha \approx 0.7$). The~major-merger rate rises mildly with stellar mass, roughly $\mathcal{R} \propto \mstar^{0.3}$~\citep{xuMajorMergerGalaxyPairs2012,lopez-sanjuanDominantRoleMergers2012,robothamGalaxyMassAssembly2014}, and~simulations place the strongest mass dependence above \mbox{$\sim$$2 \times 10^{11} \msun$~\citep{rodriguez-gomezMergerRateGalaxies2015}}. Integrating this rate over $z=1 \to 0$ gives of order $0.5$--$1$ major mergers per massive QG, enough to raise the zero point by $\sim$0.05--0.1 dex (Appendix~\ref{appsec:models}). Because~the track is steeper than the relation, it also steepens the slope mildly. This steepening is only a sub-$1\sigma$ tendency, and~its amplitude and even its sign depend on the assumed track slope and on the mass dependence of the rate. Over~the same interval, these mergers supply only $\sim$15--20\% of the observed QG size growth. Net effect: the zero point rises in both UV and optical, the~slope steepens mildly in both bands, and~the wavelength contrast is~negligible.

\subsubsection{Minor Mergers and~Accretion}\label{sssec:minorQG}

Minor mergers and smooth accretion are now widely seen as the dominant driver of QG size growth at $z \lesssim 2$~\citep{naabMinorMergersSize2009,oserTwoPhasesGalaxy2010,oserCosmologicalSizeVelocity2012,hilzHowMinorMergers2013,vandokkumGrowthMassiveGalaxies2010,newmanCanMinorMerging2012,hopkinsMergersBulgeFormation2010,rodriguez-gomezStellarMassAssembly2016,pulsoniExtendedPlanetaryNebula2018,williamsGrowthGalaxyStellar2025}. A~galaxy accreting this material climbs a steep merger track, gaining size about twice as fast as mass (\mbox{$\mathrm{d}\log\re$/$\mathrm{d}\log\mstar \approx 2$})~\citep{naabMinorMergersSize2009,hilzHowMinorMergers2013}. That is twice the slope of the equal-mass dry-merger track introduced above in Section~\ref{sssec:majorQG}.

Unlike the dry-major case, the~accreted satellites are typically less massive and, crucially, their stars are younger or more metal-poor than the old in-situ population~\citep{pillepichHaloMassAssembly2014,cooperGalacticAccretionOuter2013}. They settle preferentially in the extended outer envelope~\citep{williamsGrowthGalaxyStellar2025}. There, their young, metal-poor stars contribute a larger share of the UV light than of the stellar mass. The~QG outer envelope therefore becomes UV-brighter than the in situ core, and~the half-light radius of an individual QG grows faster in the UV than in the~optical.

The single galaxy moves up and to the right in the size--mass plane, with~a small mass change but a large size change, and~its outer envelope becomes UV-brighter than its centre. Both the accreted-to-in situ fraction and the minor-merger rate peak at high mass, above~$\log\mstar$/$\msun \simeq 10.7$~\citep{bluckStructuresTotalMinor2012,pillepichHaloMassAssembly2014,pulsoniStellarHalosETGs2021}. The~high-mass end therefore grows most, which steepens the SMR slope and raises the zero point, with both effects larger in the UV than in the optical. Net effect: the slope steepens and the zero point rises in both bands more strongly in the UV. This large UV-biased contrast is one of the cleanest empirical fingerprints of minor mergers and accretion in the QG~SMR.

\subsubsection{Adiabatic~Expansion}\label{sssec:adiabaticQG}

Adiabatic expansion shapes a QG at the moment it quenches. The~final starburst, together with any quasar-mode AGN activity, expels a large fraction of the central baryons from the potential~\citep{fanDramaticSizeEvolution2008,damjanovRedNuggets152009}. When this loss is slow compared with the local dynamical time, the~stellar orbits re-equilibrate into a less bound configuration and conserve the adiabatic invariant~\citep{hillsEffectMassLoss1980}. In~this adiabatic limit $\re$ grows in inverse proportion to the enclosed mass, and~idealized models give a factor of $\sim$2 for a mass loss $\Delta M/M \sim 0.3$--$0.5$~\citep{fanDramaticSizeEvolution2008,ragone-figueroaPuffingEarlytypeGalaxies2011}. Such large expellable gas fractions exist only during this gas-rich quenching event. Once quenched, galaxies at $z < 1$ hold low molecular-gas fractions of order a few per cent~\citep{youngATLAS3DProjectIV2011,spilkerMolecularGasContents2018}. Hence, any later gas expulsion enlarges $\re$ only~modestly.

Adiabatic expansion moves each star outward by a factor set only by the mass removed inside its orbit, regardless of its age. In~an evolved QG the ongoing mass loss comes from the material that aging stars shed, which follows the stellar distribution. Therefore, expelling it stretches the galaxy nearly uniformly, and~the UV and optical sizes grow by nearly the same factor, whatever the colour gradient. The~one exception is the central gas expelled at the quenching event. Being more centrally concentrated than the stars, the~expelled gas makes the inner regions expand more. This uneven expansion can stretch a pre-existing radial age gradient into a small difference between the UV and optical sizes. However, that difference happens while quenching (not after quenching), and its effect on the population is negligible. Adiabatic expansion therefore grows both bands by nearly the same amount, unlike the wavelength-asymmetric mergers and~stripping.

The galaxy moves nearly straight up in the size--mass plane, gaining size while losing little mass. The~two feedback channels act at opposite ends of the mass function. Quasar-mode episodes affect the massive compact progenitors, but~they are triggered only at quenching and setting the size of newly quenched galaxies~\citep{damjanovRedNuggets152009,trujilloDissectingSizeEvolution2011,settonCompactStructuresMassive2022}. Within~the established population, the~ongoing expansion is driven instead by the material still expellable, dominated at late times by the slow removal of aging stars~\citep{ragone-figueroaPuffingEarlytypeGalaxies2011,ragone-figueroaBCGMassEvolution2018,salesFeedbackStructureSimulated2010}. This fraction is largest in low-mass systems with shallow potentials. Adiabatic expansion therefore lifts the low-mass end of the QG SMR most.  Net effect: the slope flattens, the zero point rises in both bands, and~the wavelength contrast is~negligible.

\subsubsection{High-Velocity~Encounters}\label{sssec:harassmentQG}

For QGs in galaxy clusters, the~same high-velocity encounters that affect SFGs continue to operate but without the gas-driven nuclear starbursts~\citep{mooreGalaxyHarassmentEvolution1996,mooreMorphologicalTransformationGalaxy1998,mastropietroMorphologicalEvolutionDiscs2005,smithSensitivityHarassmentOrbit2015,lamarcaGalaxyPopulationsHydra2022}. Repeated tidal shocks heat the stellar distribution, strip outer material, and~increase the central concentration of the remnant. The~observational evidence comes from the early-type dwarf populations of nearby clusters such as Virgo and Hydra~I~\citep{liskerVirgoClusterEarlyType2006a,lamarcaGalaxyPopulationsHydra2022}, whose structural properties are reproduced by $N$-body high-speed encounter simulations~\citep{smithSensitivityHarassmentOrbit2015}.

The wavelength response is set by the metallicity of the removed material, because~a metal-poor outer shell stays UV-bright even when it is old (Section~\ref{sec:framework}). Both the binding energy of the outskirts and their metallicity contrast rise steeply with mass~\citep{rodriguez-gomezStellarMassAssembly2016}. A~massive QG therefore holds a tightly bound, metal-poor, ex-situ envelope, whereas a low-mass dwarf has only a loosely bound, in situ, near-solar outskirt whose removal carries no wavelength contrast in $\re$.

Above the pivot mass considered here, fast flybys heat that outer envelope but rarely unbind it, and~most orbits strip no stars at all~\citep{smithSensitivityHarassmentOrbit2015}. Its removal is instead the work of the sustained cluster tide (Section~\ref{sssec:tidalQG}), leaving high-velocity encounters with no measurable imprint on the QGs we study. The~effect is significant only below the SMR pivot point, a~mass regime absent from Figure~\ref{fig:qg}, where repeated shocks compact loosely bound dwarfs~\citep{mastropietroMorphologicalEvolutionDiscs2005,lamarcaGalaxyPopulationsHydra2022}. Net effect: above the pivot the QG slope and zero point are essentially unchanged with no UV--optical contrast, and~any effect is confined to the out-of-scope low-mass~regime.

\subsubsection{Tidal~Stripping}\label{sssec:tidalQG}

Tidal interactions in galaxy groups, clusters, and~the central regions of host haloes can remove stars from the loosely bound outer envelopes of QGs~\citep{merrittRelaxationTidalStripping1983,aguilarDensityProfilesTidally1986,gnedinTidalEffectsClusters2003,liskerVirgoClusterEarlyType2006a,blomSLUGGSSurveyNew2014,lokasTidalEvolutionGalaxies2020,mayesContributionStrippedNuclei2021,yiFormationPathwaysCompact2025,chambaImpactEnvironmentSize2024}. This envelope is the diffuse, low-surface-brightness ex situ stellar halo at large radius (roughly beyond $\sim$$2~\re$), structurally distinct from the compact in situ core. Like high-velocity encounters, tidal stripping preferentially removes this outermost material, but~the two differ in dynamical regime. The~former proceeds through impulsive flyby encounters that shock-heat the stellar distribution at all radii, whereas the latter acts adiabatically through the smooth host potential and leaves the inner stellar core relatively undisturbed. 
That structural distinction sets which stars are removed, while the wavelength contrast developed below follows from the metallicity gradient of the removed material rather than from the impulsive or adiabatic character of the~removal.

The outer envelope of a massive QG is built by past minor mergers~\citep{pulsoniStellarHalosETGs2021,williamsGrowthGalaxyStellar2025}. As~we mention in Section~\ref{sssec:minorQG}, this accreted material is more metal-poor than the in situ core at the same old age, and~therefore UV-brighter~\citep{pillepichHaloMassAssembly2014,cooperGalacticAccretionOuter2013,georgeTwoRestframeWavelength2024}. Because~it shines disproportionately in the UV, adding or removing it changes the UV size more than the optical. Minor mergers build this envelope in the field galaxies and raise their UV size, whereas tidal stripping removes it from cluster galaxies and shrinks their UV size more than the optical~one.

A tidally stripped QG moves down and to the left, but~not along the relation. Removing the outer envelope shrinks the $\re$ faster than $\mstar$, giving a steep track ($\mathrm{d}\log\re/\mathrm{d}\log\mstar > 1$) well above $\alpha \approx 0.7$, and~the galaxy drops below the SMR rather than sliding along it. Stripping effects are strongest at the lower-mass end of the QG range, where satellite fractions are higher and potentials shallower~\citep{smithPreferentialTidalStripping2016,rheePhasespaceAnalysisGroup2017,mayesContributionStrippedNuclei2021}. That end therefore drops furthest, and~both slopes steepen. Because~the removed envelope is metal-poor and UV-bright, its loss drops the UV $\re$ more than the optical, and~the UV slope steepens more. The~per-galaxy deficit is large, with~cluster QGs observed being smaller than the field~\citep{chambaImpactEnvironmentSize2024,georgeEffectsEnvironmentSize2025}, though~only a minority are deeply stripped, leaving the population amplitude small. Net effect: Both slopes steepen and both zero points fall, more strongly in the UV than in the optical. 
This is the~time-reverse of minor-merger growth, which raises the UV zero point faster than the~optical. 

\subsubsection{Progenitor~Bias}\label{sssec:progenitor-bias}

Progenitor bias is a population-level effect rather than a physical process on individual galaxies. When an SFG quenches, it joins the QG population as a newcomer. At~fixed stellar mass, that newcomer is larger than the QGs which quenched earlier and have since compacted through mergers and stripping~\citep{vandokkumMorphologicalEvolutionAges2001,sagliaFundamentalPlaneEDisCS2010,carolloNewlyQuenchedGalaxies2013,cassataConstrainingAssemblyNormal2013,belliStellarPopulationsSpectroscopy2015,fagioliMinorMergersProgenitor2016,genelSizeEvolutionStarforming2018,damjanovSizeSpectroscopicEvolution2023,georgeTwoRestframeWavelength2024,estrada-carpenterCLEARMorphologicalEvolution2023}. The~relation shifts not because any galaxy grows or gains mass, but~because the size distribution at fixed mass steadily gains larger members, raising the mean $\re$ and lifting the zero point. This zero point offset survives a $\sim$10--30\% disk-fading correction and a mild bias of quenching systems toward the compact side of the star-forming size distribution. Most galaxies that quench are of modest mass ($\log\mstar/\msun \sim 10$--$10.7$), simply because low-mass galaxies far outnumber high-mass ones~\citep{muzzinEvolutionStellarMass2013,tomczakGalaxyStellarMass2014,davidzonCOSMOS2015GalaxyStellar2017}, even though any single massive galaxy is more likely to quench~\citep{pengMassEnvironmentDrivers2010,bluckBulgeMassKing2014,schreiberHerschelViewDominant2015}.  {This picture applies more to the secular, non-merger quenching that dominates at these masses, whereas merger-driven quenching instead compacts the newcomer and is already captured by the merger channels above.}

A newcomer arrives with an extended young disk~\citep{belliMOSFIRESpectroscopyQuiescent2019,suessColorGradientsQuiescent2020,yanoRelationGalaxyStructure2016,estrada-carpenterCLEARMorphologicalEvolution2023},  {often bearing the transitional morphology of the green valley, where quenching disks show a surplus of rings and lenses~\citep{kelvinGalaxyMassAssembly2018,smithGalaxyMassAssembly2022}}. Because~its young stars are UV-bright, this disk makes the newcomer's UV size larger than its optical size at first. The~disk then fades from the UV within $\sim$1 Gyr, set by the lifetimes of A and late-B main-sequence stars, but~keeps shining in the optical for several Gyr. A~newcomer's UV size therefore shrinks more quickly than its optical size, and~the optical zero point of the QG SMR accordingly receives a larger boost than the UV zero point. This optical-over-UV contrast is modest, and~it weakens toward higher redshift. There, rapid recent quenching leaves many QGs still inside their $\sim$1 Gyr UV-bright window, and~their surviving UV disks increase the UV zero point on par with the optical zero~point.

The newcomer influx peaks at modest masses ($\log\mstar/\msun \sim 10$--$10.7$), and~their continuous addition increases the optical zero point of the QG SMR. Because~they arrive preferentially near and below the pivot point, they raise the low-mass sizes more than the massive end and therefore flatten $\alpha$. The~UV zero point also rises but with a smaller amplitude because~each newcomer's UV light fades within $\sim$1 Gyr while its optical light persists. Net effect: the optical zero point rises more than the UV, and~both slopes flatten. At~$z \lesssim 1$ the newcomer flux multiplies the low-mass QG number density by~a factor of $\sim$2--3 at $\log\mstar/\msun \approx 10$ while barely affecting the massive end~\citep{moutardVIPERSMultiLambdaSurvey2016,davidzonCOSMOS2015GalaxyStellar2017}. Progenitor bias therefore reshapes the low-mass slope, while minor mergers steepen the massive end, and~the observed slope is the net of the two. Progenitor bias, adiabatic expansion, and~minor-merger accretion all raise the QG intercept, and~their UV--optical contrasts, optical-strong, neutral, and~UV-strong, are what separate the~three.

\section{Caveats and~Degeneracies}\label{sec:caveats}

The idealized wavelength contrasts predicted in Section~\ref{sec:processes} are subject to several observational systematics that compete with the process-driven signals in amplitude and limit the quantitative interpretation of any UV+optical SMR measurement. Below~we flag, in~order, the~stellar-population caveats (age--metallicity degeneracy, dust attenuation, UV upturn, AGN contamination), the~structural-modelling caveats (single-S\'ersic vs.\ bulge+disk decompositions), and~the measurement-systematics caveats (baseline colour gradients, PSF/resolution mismatch, surface-brightness dimming).

\textit{Age--metallicity degeneracy.} Both stellar age and stellar metallicity affect the strength of the $4000$ \AA\ break and the slope of the rest-UV continuum~\citep{wortheyComprehensiveStellarPopulation1994,tangDistinguishingAgeMetallicity2013,meulenaerDerivingPhysicalParameters2014}. A~metal-poor old population can mimic the UV brightness of a younger metal-rich population and therefore inflate the apparent UV size of an old galaxy. The~cleanest discriminant is spatially resolved spectroscopy from instruments such as MUSE, ERIS, NIRSpec IFU, and~GIRMOS~\citep{urrutiaMUSEWideSurveySurvey2019,fosterMAGPISurveyScience2021,daviesERISRevitalisingAdaptive2018,bokerNearInfraredSpectrographNIRSpec2022,jakobsenNearInfraredSpectrographNIRSpec2022,conodAdaptiveOpticsSystem2023,hayozHighcontrastSpectroscopyNew2025}, which allows direct measurement of stellar-population gradients. In~practice, this route is limited by the low surface brightness of galaxy outskirts, which is exactly where the population gradients matter most, and~it is cleanest for galaxies whose outer regions stay bright enough to reach adequate signal-to-noise ratio per radial~bin.

\textls[-20]\textit{Dust attenuation.} Dust reddens galaxies unevenly across their light profiles \mbox{\citep{calzettiDustContentOpacity2000,salimDustAttenuationCurves2018,salimDustAttenuationLaw2020,nelsonSpatiallyResolvedDust2016,linALMaQUESTIVALMAMaNGA2020,magnelliCEERSMIRIDeciphers2023,tadakiSpatialExtentMolecular2023,liEstimatingDustAttenuation2024}.}  {Because dust is centrally concentrated, it suppresses the central UV light and biases UV size measurements upward mimicking intrinsic inside-out growth~\citep{nelsonSpatiallyResolvedDust2016,nedkovaUVCANDELSRoleDust2024}}. The~bias is largest for the most strongly star-forming SFGs and for wet-merger remnants. Dust content is small but not negligible in transitional and recently quenched QGs (post-starburst, K+A, fading-disk newcomers; Section~\ref{sssec:progenitor-bias}), where residual ISM and centrally concentrated dust lanes can still bias UV size measurements during the first $\sim$1 Gyr after quenching. Resolved attenuation maps from JWST/MIRI and ALMA, combined with energy-balance SED fitting, are well suited to correct for dust effects. A~spatially resolved multi-band UV-optical-NIR SED fit, performed pixel-by-pixel or in radial annuli, can also recover the attenuation as a function of radius and correct the inferred UV sizes. The~mid-IR and sub-mm data are ideal rather than~essential.

\textit{UV upturn in old populations.} Hot extreme-horizontal-branch stars and post-AGB stars in the most massive, oldest QGs produce a ``UV upturn'' that contributes non-negligible flux at $\lambda_{\rm rest} \lesssim 2500$ \AA~\citep{codeUltravioletPhotometryOrbiting1979,greggioCluesHotStar1990,yiGalaxyEvolutionExplorer2005,lecrasModellingUVSpectrum2016,lonoceStellarPopulationProperties2020,akhilDecipheringPropertiesUV2024,martocchiaVirgoEnvironmentalSurvey2025}. The~rest-frame near-UV at $\sim$3000 \AA\ adopted here lies longward of the upturn and avoids most of this contamination. A~residual concern is a broad-band filter that at certain redshifts still samples $\lambda_{\rm rest} \lesssim 2500$ \AA, which can mimic extended UV light from a young population and is worst in the most massive QGs at $z \lesssim 0.5$~\citep{lecrasModellingUVSpectrum2016}.

\textit{AGN contamination.} Unresolved nuclear UV point sources from low-luminosity AGN inflate the inferred central S\'ersic concentration and bias $\re(\textrm{UV})$ downward relative to $\re(\textrm{opt})$, mimicking the compact-nuclear-starburst signature attributed to wet major mergers (Section~\ref{sssec:wet-mergers}) and the post-quasar phase invoked in adiabatic expansion. Multi-wavelength AGN identification (X-ray, mid-IR, radio) is required to flag affected systems before~fitting.

\textit{Single-S\'{e}rsic versus bulge+disk biases.} Many of the predictions above assume that the global $\re$ measured by a single-S\'{e}rsic fit responds to bulge growth and disk evolution together. A~growing body of evidence shows that the global $\re$ can deviate substantially from the disk $\re$ when the bulge becomes structurally distinct~\citep{davariDetectionProminentStellar2017,moslehGalaxySizes22020,robothamProFusePhysicalMultiband2022,kawinwanichakijStellarMassSizeRelation2025}. The~clearest test of the predictions in Table~\ref{tab:summary} is therefore at the level of bulge+disk decompositions, not single-band single-component fits. However, such a decomposition carries roughly twice the free parameters of a single-S\'{e}rsic fit, and~the bulge and disk become degenerate at low signal-to-noise and coarse resolution. Reliable decompositions are therefore feasible only for bright, well-resolved galaxies, while fainter and smaller systems remain limited to single-S\'{e}rsic~sizes.

\textit{Baseline colour gradients.} Galaxies have negative colour gradients (bluer outskirts) at essentially all masses and redshifts~\citep{labarberaSPIDERSampleGalaxy2010,moslehConnectionStellarMass2017,dimauroStructuralPropertiesClassical2019,suessHalfmassRadii70002019,suessDissectingSizeMassS1Mass2021,casuraGalaxyMassAssembly2022,nedkovaBulge+discDecompositionHFF2024}, which generically produces $\re(\textrm{UV}) > \re(\textrm{opt})$ regardless of which of the twelve processes above is at work. The~predictions in Table~\ref{tab:summary} are therefore \emph{differential} signals (changes in the wavelength contrast) on top of this universal baseline. The~absolute UV-vs-optical size comparisons cannot be interpreted process-by-process without first subtracting it. Because~this baseline is nearly a constant offset, subtracting it preserves the process-induced signals rather than erasing them, though~how cleanly each is recovered depends on its amplitude relative to the residual baseline scatter. The~near-neutral processes, most notably adiabatic expansion, carry the smallest differential contrast and are the hardest to recover against the residual baseline scatter. The~strong-contrast channels, such as minor-merger accretion and progenitor bias, remain clearly separable above~it.

\textit{PSF and resolution.} The angular resolution (PSF) of both ground-based and space-based imaging is wavelength-dependent. This effect biases UV-vs-optical size differences in ways that depend on intrinsic profile shape and S/N. Although~forward-modelled, PSF-matched single-S\'ersic fits or PSF-convolved bulge+disk decompositions can mitigate this issue, extensive simulations are required to quantify the biases in the size measurements~\citep{georgeTwoRestframeWavelength2024,georgeEffectsEnvironmentSize2025}.

\textit{Surface-brightness limits.} Although cosmological $(1+z)^4$ dimming is purely geometric and wavelength-independent~\citep{tolmanEstimationDistancesCurved1930},  the~recovered $\re$ is not. Magnitude-limited imaging imposes limits on the surface brightness of the outer profile regions that can be traced. These surface brightness limits, especially at the outskirts, can have a negative impact on the measured sizes. 
Since the intrinsic rest-UV profile is typically more extended than the optical~\citep{vanderwel3DHST+CANDELSEvolutionGalaxy2014,georgeTwoRestframeWavelength2024,georgeEffectsEnvironmentSize2025}, UV sizes have a stronger negative bias than the optical sizes. This wavelength-dependent bias partly masks the intrinsic $\re(\textrm{UV}) > \re(\textrm{opt})$ contrast. These biases can be quantified and corrected by the same forward-modelling injection-recovery simulations used for the PSF caveat above, run at the effective surface-brightness limit of each~band.


\section{Summary, Diagnostics, and~Outlook}\label{sec:summary}

 {We have reviewed, process by process, the~response of the galaxy \smr\ in rest-frame UV and rest-frame optical, separately for SFGs and QGs, and~set out the expected qualitative direction of each response, compiled in Table~\ref{tab:summary} and visualized in Figures~\ref{fig:sfg} and \ref{fig:qg}.}

The central observation is that a comparison of the UV \smr\ with the optical \smr\ lifts several of the degeneracies that affect single-band analyses. Three cases deserve emphasis here. First, in~QGs the three processes that raise the zero point the SMR---minor mergers and accretion, adiabatic expansion, and~progenitor bias---predict different wavelength contrasts: minor-merger growth raises the UV intercept \emph{more} than the optical, adiabatic expansion raises both nearly equally (the near-neutral case hardest to separate from the baseline colour gradient; Section~\ref{sec:caveats}), and~progenitor bias raises the optical intercept slightly more than the UV. The~relative ranking of UV and optical zero-point evolution at fixed $\mstar$, combined with the sign of the slope change (steepening for minor mergers, flattening for the other two), therefore directly diagnoses the dominant channel\footnote{Dry major mergers also raise the SMR intercept with a mild slope steepening tendency. However, this change is much lower than that from minor mergers and accretion. Additionally, due to lack of its wavelength contrast, the~major merger-driven change is significantly different from minor-merger accretion.}. Second, in~SFGs the apparent deceleration of size growth at $z \lesssim 1$~\citep{vanderwel3DHST+CANDELSEvolutionGalaxy2014,georgeTwoRestframeWavelength2024,kawinwanichakijStellarMassSizeRelation2025} can be ascribed either to a pause in inside-out growth or to bulge-driven contraction of the global $\re$. These two processes yield opposite UV--optical contrasts in SFG SMR evolution (bulge growth lowers the optical $\re$ more than the UV; inside-out growth raises the UV $\re$ at least as fast as the optical) and are therefore empirically separable in surveys with both rest-UV and rest-optical imaging. Third, environmental processes (harassment, ram-pressure and tidal stripping) leave a wavelength fingerprint biased to the UV side in SFGs but much weaker in QGs---where only tidal stripping of the UV-bright accreted envelope retains a UV bias, harassment being nearly wavelength-neutral---providing an independent test of cluster-vs-field SMR comparisons~\citep{matharuHSTWFC3Grism2019,chanKMOSClusterSurvey2018,afanasievGalaxyMasssizeRelation2023,georgeEffectsEnvironmentSize2025}.

These considerations translate into four specific, near-term diagnostics:

\begin{enumerate}
\item {The sign of $\alpha_{\rm UV}-\alpha_{\rm opt}$ for SFGs at $z \lesssim 1$ discriminates inside-out growth from bulge-driven inside-out quenching.} Bulge growth decreases the optical $\re$ at high $\mstar$ and flattens the optical slope while leaving the UV slope nearly unchanged, driving $\alpha_{\rm UV} - \alpha_{\rm opt}$ positive; pure inside-out growth flattens the two slopes by comparable amounts, keeping $\alpha_{\rm UV} - \alpha_{\rm opt} \approx 0$ while increasing the UV zero point. A~significantly positive $\alpha_{\rm UV} - \alpha_{\rm opt}$ at fixed $z$ therefore signals bulge-driven quenching, whereas a null slope difference accompanied by a rising UV intercept favours disk-dominated inside-out growth. Extending the joint UV+optical \smr\ measurements from CLAUDS+HSC~\citep{georgeTwoRestframeWavelength2024} at $0.1 < z < 0.9$ to higher redshifts with LSST, Euclid, Roman and JWST/NIRCam will map the cosmic epochs at which each channel dominated SFG structural evolution, and~in particular when bulge growth begins to compete with disk~assembly.

\item {Bulge growth decelerates the SFG \emph{global} (single-S\'{e}rsic) zero-point evolution in the optical relative to the UV, whereas the \emph{disk} zero point evolves at comparable rates in both bands, tilted only mildly UV-ward by inside-out growth.} The two channels (bulge growth shrinking the optical global $\re$; inside-out growth increasing the UV disk $\re$) act in opposite directions on the two measurements. The~UV-versus-optical split in the global single-S\'{e}rsic SMR slope can flag which channel dominates, because~bulge growth flattens the optical slope while leaving the UV slope nearly unchanged, whereas inside-out growth flattens both alike. However, two-wavelength bulge--disk decompositions~(e.g.,~\citep{robothamProFusePhysicalMultiband2022,kawinwanichakijStellarMassSizeRelation2025}) are then needed to measure the bulge and disk zero points separately to confirm the findings of single-S\'{e}rsic~modelling.

\item {Cluster--field SMR differences are larger in UV than in optical, with~the contrast more pronounced in SFGs than in QGs.} Ram-pressure stripping (SFGs only) and tidal stripping (both populations) preferentially remove UV-bright outer material---gas-rich disks in SFGs and ex situ accreted envelopes in QGs~\citep{mooreGalaxyHarassmentEvolution1996,smithSensitivityHarassmentOrbit2015,boselliRamPressureStripping2022}. We therefore predict $|R_{e,\rm UV}^{\rm cluster} - R_{e,\rm UV}^{\rm field}|  >  |R_{e,\rm opt}^{\rm cluster} - R_{e,\rm opt}^{\rm field}|$ for both populations at fixed $\mstar$, testable at $z \lesssim 1$ with existing rich-cluster and field surveys~(e.g.,~\citep{matharuHSTWFC3Grism2019,georgeTwoRestframeWavelength2024,georgeEffectsEnvironmentSize2025}), unless~preprocessing is more important than cluster-specific processes. This prediction is also testable at $z \gtrsim 1$, treated as a distinct regime, with~JWST/NIRCam plus Roman, where the high-redshift (proto)cluster environments host a different member population and mass function and must be compared at fixed redshift and halo~mass.

\item {The intrinsic scatter of the UV SMR exceeds that of the optical SMR for both SFGs and QGs.} Several of the processes that show wavelength contrast in SMR evolution (mergers, stripping, in situ growth) produce stochastic, transient UV-bright outliers even when their effect on the median \smr\ is modest. Initial evidence from CLAUDS+HSC over the COSMOS field at $0.1 < z < 0.9$~\citep{georgeTwoRestframeWavelength2024} already shows the UV scatter exceeding the optical; extending the measurement to the wider area and higher redshifts probed by LSST, Euclid, Roman and JWST/NIRCam will test whether the same wavelength asymmetry holds across cosmic time. 
\end{enumerate}

{The diagnostic predictions here are qualitative and directional rather than numerical, but~they are testable in exactly this differential form. These diagnostics (the sign of $\alpha_{\rm UV}-\alpha_{\rm opt}$, the~cluster--field UV-versus-optical size contrast, and~the excess UV scatter) can each be measured with joint UV+optical imaging. The~per-process prescriptions of Appendix~\ref{appsec:models} are equally portable and~can be incorporated into semi-analytic and semi-empirical models of galaxy evolution to turn these qualitative directions into quantitative, model-specific~predictions.}

The unifying message is that the joint UV+optical \smr\ is no longer a niche measurement: with CLAUDS+HSC, LSST, Euclid, and~Roman now or soon delivering rest-frame UV and optical imaging out to $z \sim 2$ for samples of $10^{6}$--$10^{7}$ galaxies, two-rest-frame-wavelength SMR analysis can become a standard observable. The~addition of rest-frame near-IR sizes from Euclid NISP and Roman WFI will extend this to a three-wavelength view in which the near-IR anchors the bulk-mass radius even more robustly than the optical. Confronting it with the process-by-process predictions of Table~\ref{tab:summary} should allow the community to identify which structural-evolution channels dominate at which mass and redshift, and~to use the residuals as a direct test of the sub-grid feedback prescriptions in cosmological~simulations.

\vspace{6pt}
\authorcontributions{Conceptualization, A.G., M.S. and I.D.; methodology, A.G.; investigation, A.G.; writing---original draft preparation, A.G.; writing---review and editing, A.G., M.S. and I.D.; visualization, A.G.; supervision, M.S. and I.D.; funding acquisition, M.S. and I.D. All authors have read and agreed to the published version of the~manuscript.}

\funding{This research was supported by the Institute of Astronomy and Astrophysics, Academia Sinica (ASIAA), and~by Saint Mary's University, Halifax. M.S.\ and I.D.\ acknowledge support from Discovery Grants (RGPIN-2018-05425, DDG-2024-00004, RGPIN-2020-06023, and RGPAS-2020-00065) from the Natural Sciences and Engineering Research Council (NSERC) of Canada.}



\dataavailability{No new observational data were created or analysed in this study. The~theoretical modelling frameworks and synthesized results are fully presented within the article and its appendix.  
}

\acknowledgments{We thank the anonymous referees for their comments that helped us improve the quality of this manuscript. This synthesis grew out of the doctoral research of A.G.\ under the supervision of I.D.\ and M.S.\ at Saint Mary's University, and~builds on data products from the CFHT Large-Area U-band Deep Survey (CLAUDS;~\citep{sawickiCFHTLargeArea2019}) and the Subaru Hyper Suprime-Cam Strategic Program (HSC-SSP;~\citep{aiharaHyperSuprimeCamSSP2018,aiharaThirdDataRelease2022}). This work is based in part on observations obtained at the Canada--France--Hawaii Telescope (MegaCam) and at the Subaru Telescope (HSC). The~analysis made use of NumPy~1.26.4~\citep{harrisArrayProgrammingNumPy2020}, SciPy~1.15.2~\citep{virtanenSciPy10Fundamental2020}, and~Matplotlib~3.10.8~\citep{hunterMatplotlib2DGraphics2007}. During~the preparation of this manuscript, the~authors used AI for language editing, figure-code drafting, and~literature cross-checking, and~take full responsibility for all reviewed and edited~output.}

\conflictsofinterest{The authors declare no conflicts of interest. The~funders had no role in the design of the study; in the collection, analyses, or~interpretation of data; in the writing of the manuscript; or in the decision to publish the~results.}


\abbreviations{Abbreviations}{%
~The following abbreviations are used in this manuscript:\\

\noindent
\begin{tabular}{@{}ll}
SMR     & Size--mass relation \\
SFG     & Star-forming galaxy \\
QG      & Quiescent galaxy    \\
UV      & Ultraviolet         \\
SF      & Star formation      \\
SFR     & Star formation rate \\
sSFR    & Specific star formation rate \\
SFMS~~~~~~    & Star-forming main sequence \\
IFS     & Integral field spectroscopy \\
ICM     & Intracluster medium \\
AGN     & Active galactic nucleus \\
\end{tabular}

\noindent
\begin{tabular}{@{}ll}
SN      & Supernova \\
IMF     & Initial mass function \\
ORCID   & Open Researcher and Contributor ID \\
NSERC   & Natural Sciences and Engineering Research Council (Canada) \\
ASIAA   & Institute of Astronomy and Astrophysics, Academia Sinica \\
CFHT    & Canada–France–Hawaii Telescope \\
CLAUDS  & CFHT Large-Area U-band Deep Survey \\
HSC-SSP & Hyper Suprime-Cam Subaru Strategic Program \\
LSST    & Legacy Survey of Space and Time \\
\end{tabular}}

\appendixtitles{yes} 
\appendixstart
\appendix

\section{Toy Models Used to Test Each~Process}\label{appsec:models}

\subsection{Forward Monte Carlo~Framework}\label{appsec:overview}

To determine the relative strength of the impact of the processes considered in Section~\ref{sec:processes} on the SMR, we build a common forward model of the SMR and apply each process to it. 
We first seed a mock population on a power-law baseline,
\begin{equation}
\log\re = \log R_0 + \alpha (\log\mstar - \log M_0) + \mathcal{N}(0,\sigma),
\label{eq:seed}
\end{equation}
where $\re$ is the half-light radius, $\mstar$ the stellar mass, $\alpha$ the slope, and~$\sigma$ the intrinsic (Gaussian) scatter in dex. The~normalization $\log R_0$ is the characteristic size, defined at the fiducial mass $M_0=5\times10^{10} \msun$ ($\log M_0=10.7$, the~knee of the \citet{davidzonCOSMOS2015GalaxyStellar2017} mass function), and~it is at $M_0$ that we report every zero-point shift. The~SFGs follow a single slope $\alpha=0.22$ with $\log R_0=0.74$ and $\sigma=0.20$ dex. The~QGs follow a continuous broken power law whose slope changes from $\alpha=0.70$ to $\alpha\approx0.15$ at the pivot mass $\log M_{\rm p}\approx10.4$, with~$\log R_0=0.55$ and $\sigma=0.16$ dex \citep{vanderwel3DHST+CANDELSEvolutionGalaxy2014,kawinwanichakijHyperSuprimeCamSubaru2021,georgeTwoRestframeWavelength2024}. Stellar masses are drawn by inverse-cumulative sampling of the \citet{davidzonCOSMOS2015GalaxyStellar2017} double-Schechter~functions.

Each process then displaces every affected galaxy along its own size--mass track,
\begin{equation}
\Delta\log\re = s \Delta\log\mstar, \qquad s \equiv \frac{{\rm d}\log\re}{{\rm d}\log\mstar},
\label{eq:track}
\end{equation}
where $s$ is the track slope and $\Delta\log\mstar$ the change in stellar mass. Writing $\mu(\mstar)$ for the mean change in $\log\mstar$ of galaxies that start at mass $\mstar$, each galaxy ends at a mass $\log\mstar+\mu$ and a size $\log\re+s\mu$. Differentiating the new size with respect to the new mass gives a refitted slope $\alpha_{\rm new}=(\alpha+s'\mu+s\mu')/(1+\mu')$. For~the zero point, a~galaxy that ends at $M_0$ began at $\log M_0-\mu$ and has therefore gained $(s-\alpha)\mu$ in size relative to the original relation. Hence, to~first order,
\begin{equation}
\Delta\alpha \simeq \frac{s' \mu + (s-\alpha) \mu'}{1+\mu'}, \qquad
\Delta\log R_0 \simeq (s-\alpha) \mu(M_0),
\label{eq:refit}
\end{equation}
where $\Delta\alpha$ is the change in slope, $\Delta\log R_0$ the change in the characteristic size at $M_0$, and~$\mu' \equiv {\rm d}\mu/{\rm d}\log\mstar$ and $s' \equiv {\rm d}s/{\rm d}\log\mstar$ are the mass gradients of the growth and of the track slope. The~adopted track slopes follow the virial arguments of \mbox{\citet{naabMinorMergersSize2009}} and \citet{bezansonRelationCompactQuiescent2009}. For~a track whose slope does not vary with mass ($s'=0$), the~slope change reduces to $(s-\alpha) \mu'/(1+\mu')$. We adopt a single representative slope near the fiducial mass for every channel except wet major mergers, where the dissipative compaction makes $s$ explicitly mass-dependent (Appendix~\ref{appsec:notes}). Equation~(\ref{eq:refit}) therefore serves as an order-of-magnitude cross-check on the Monte Carlo rather than an exact~result.

Equation~(\ref{eq:refit}) serves as an analytic cross-check of the full Monte Carlo, in~which discrete channels (mergers, stripping) are Poisson-sampled over $z=1 \to 0$, smooth channels (in situ growth, stellar mass return) are applied continuously by integrating their rates over the same interval, and~$\mu(\mstar)$ is built from the measured event rate, per-event mass change and mass-dependent affected fraction \citep{xuMajorMergerGalaxyPairs2012,lotzMajorMinorGalaxy2011,speagleHighlyConsistentFramework2014}. We assign each galaxy a rest-UV ($3000$ \AA) and a rest-optical ($5000$ \AA) half-light radius from its stellar light in one of two ways. The~first applies an evolving colour-gradient law that sets how much larger the UV size is than the optical~\citep{vanderwel3DHST+CANDELSEvolutionGalaxy2014}. The~second treats the galaxy as its separate stellar components (bulge, disk, envelope) and gives each an age-dependent brightness in each band from single-population fading curves (piecewise power-law fits to the luminosity-per-unit-mass of \citet{bruzualStellarPopulationSynthesis2003} and FSPS~\citep{conroyPropagationUncertaintiesStellar2009}). Thus, young components are UV-bright and old ones are UV-faint. Then we measure $\re$ in each band. Either way, we fit the relation in Equation~(\ref{eq:seed}) separately to the optical size and UV~size.

Two channels need their own displacement rule. When a central mass fraction $f$ is removed, the~system re-equilibrates to a larger size $R_f$ from its initial size $R_i$. In~the slow (adiabatic) limit,
\begin{equation}
\frac{R_f}{R_i}=\frac{1}{1-f},
\label{eq:adiab}
\end{equation}
and in the fast (impulsive) limit $R_f/R_i=(1-f)/(1-2f)$, which diverges as $f \to \tfrac12$~\mbox{\citep{hillsEffectMassLoss1980,ragone-figueroaPuffingEarlytypeGalaxies2011}}. A~fast tidal encounter is instead governed by the ratio of the injected energy $\Delta E$ to the target's binding energy $|E|$,
\begin{equation}
\frac{\Delta E}{|E|}\sim\left(\frac{M_{\rm pert}}{m}\right)^{2}\left(\frac{r_h}{b}\right)^{4}\left(\frac{\sigma_v}{V}\right)^{2}\propto\frac{r_h^{3}}{m},
\label{eq:impulse}
\end{equation}
for a perturber of mass $M_{\rm pert}$ passing a target galaxy of mass $m$, half-mass radius $r_h$, and~internal velocity dispersion $\sigma_v$, at~impact parameter $b$ and relative speed $V$, using the virial scaling $\sigma_v^{2} \propto m/r_h$ \citep{spitzerDisruptionGalacticClusters1958,binneyGalacticDynamicsSecond2008,mooreGalaxyHarassmentEvolution1996,mooreMorphologicalTransformationGalaxy1998}. The~final proportionality $r_h^{3}/m\propto1/\bar\rho$, with~$\bar\rho$ the mean density, shows that diffuse low-mass systems are the most~susceptible.

These are deliberately simplified forward models meant to check the sign and rough amplitude of each imprint, validated against the sign conventions of Table~\ref{tab:summary}, not precise quantitative predictions. All toy UV profiles are attenuation-free. Therefore, the~quoted UV--optical contrasts are intrinsic stellar-population contrasts and exclude dust~gradients.

\subsection{Adopted Tracks, Affected Fractions, and~Wavelength~Splits}\label{appsec:table}

Table~\ref{tab:toymodels} lists, for~each process, the~adopted track slope or displacement rule, the~affected mass range and mean log-mass growth $\mu(M_0)$ at the fiducial mass, the~assumption that splits the rest-UV from the rest-optical size, and~the recovered imprint. The~signs reproduce Table~\ref{tab:summary} in every~case.

\subsection{Additional Notes on Individual~Processes}\label{appsec:notes}

We provide additional notes only for the processes whose toy-model amplitude estimates (Table~\ref{tab:toymodels}) depend on a specific adopted parameter or formula that needs justification. The~remaining channels (in situ growth, bulge growth, dry major mergers, and~ram-pressure stripping) follow straightforward track slopes and affected fractions~there.

\textit{Wet major mergers.} The collisionless remnant grows along the dissipationless track $s=1$~\citep{naabMinorMergersSize2009,hopkinsDiscriminatingPhysicalProcesses2010}. In~a gas-rich merger the cold gas dissipates, sinks to the centre, and~forms a compact nuclear starburst, and~the remnant ends up smaller than the dissipationless expectation~\citep{covingtonPredictingPropertiesRemnants2008,covingtonRoleDissipationScaling2011,hopkinsDissipationExtraLight2009}. We capture this with a per-event displacement $\Delta\log\re=\log(1+\mu_{\rm r})-\log(1+f_{\rm gas}/f_0)$, where $\mu_{\rm r}$ is the merger mass ratio, for~which we adopt a mean of $0.5$ and hence a per-event growth $\mu_{\rm ev}=0.18$~dex, and~$f_{\rm gas}$ the cold-gas fraction at coalescence~\citep{tacconiPHIBSSUnifiedScaling2018}. The~compaction scale $f_0=0.27$ is adopted, within~the range $0.25$--$0.30$ that the dissipation models motivate~\citep{covingtonPredictingPropertiesRemnants2008,covingtonRoleDissipationScaling2011,hopkinsDissipationExtraLight2009}. The~effective slope $s_{\rm eff}=1-\log(1+f_{\rm gas}/f_0)/\log(1+\mu_{\rm r})$ falls with gas fraction and turns negative for gas-rich low-mass mergers. Because~$f_{\rm gas}$ declines with stellar mass, $s_{\rm eff}$ rises with mass, and~this is the one channel for which we propagate a mass-dependent track slope, that is $s'\neq0$ in Equation~(\ref{eq:refit}). During~the starburst the young light is centrally concentrated, giving a transient per-event UV size drop of $\approx-0.25$~dex relative to the optical~\citep{wuytsSizesKinematicsGradients2010}. The~merger rate meanwhile rises weakly with mass as $\mstar^{0.3}$~\citep{xuMajorMergerGalaxyPairs2012}, giving $\mu'>0$, and~the two gradients enter $\Delta\alpha$ with opposite signs and largely cancel. Massive galaxies are moreover gas-poor, hence the population-level imprint stays small, with~$\Delta\alpha\approx0$ and $|\Delta\log R_0|\lesssim0.12$~dex.


\begin{table}[H]
\caption{Toy-model ingredients and recovered imprints for the twelve processes. The~track slope is $s\equiv{\rm d}\log\re/{\rm d}\log\mstar$, and~$\mu(M_0)$ is the mean log-mass growth (dex) at the fiducial mass $M_0=5\times10^{10} \msun$. The~recovered $\Delta\alpha$ and $\Delta\log R_0$ (dex) are over $z=1 \to 0$, optical unless noted, with~the amplitude grade of Table~\ref{tab:summary} in~parentheses.\label{tab:toymodels}}
\begin{adjustwidth}{-\extralength}{0cm}
\small
\begin{tabularx}{\fulllength}{@{}>{\raggedright\arraybackslash}X >{\raggedright\arraybackslash}X >{\raggedright\arraybackslash}X >{\raggedright\arraybackslash}X >{\raggedright\arraybackslash}X@{}}
\toprule
\textbf{Process} & \textbf{Track $\boldsymbol{s}$/Rule} & \textbf{Affected Range, $\boldsymbol{\mu}$} & \textbf{Wavelength Split} & \textbf{Recovered Imprint} \\
\midrule
\multicolumn{5}{@{}l}{\textit{Star-forming galaxies}}\\ \midrule
In situ SF, inside-out growth & $s=0.30$ & whole pop., $\mu=0.35$ & evolving colour gradient, UV zero point $\sim$+0.01 over opt & $\Delta\alpha=-0.04$, $\Delta\log R_0=+0.04$ (significant) \\ \midrule
Bulge growth, inside-out quenching & light-weighted shrink at fixed $\mstar$, $R_{\rm b}/R_{\rm d}=0.2$ & $\log\mstar>10$ SFGs & disk/bulge $L/M$: 1.5 (opt), 20 (UV) & $\Delta\alpha=-0.04$, $\Delta\log R_0=-0.03$, UV$\approx0$ (modest) \\ \midrule
Wet major mergers & $s=1$ less compaction $\log(1+f_{\rm gas}/f_0)$, $f_0=0.27$ & $0.5$--$1.3$ events, $\mu_{\rm ev}=0.18$ (per event) & per-event UV compaction $-0.25$ (burst) & $\Delta\alpha\approx0$, $|\Delta\log R_0|\lesssim-0.12$ (modest) \\ \midrule
Minor mergers, accretion & $s=1$ ($0.5$--$2$) & ex situ $\mu(M_0)=0.03$ & shared track, UV$\sim$opt & $\Delta\alpha=+0.03$, $\Delta\log R_0=+0.03$ (modest) \\ \midrule
High-velocity encounters & $s=2$ (outer-disk stripping) & $f_{\rm pop}\approx0.004$, $\log\mstar<10$ & UV outer disk first, per-event $-0.10$ & $\Delta\alpha,\Delta\log R_0\lesssim10^{-3}$ (negligible) \\ \midrule
Ram-pressure, tidal stripping & UV near-vertical $s_{\rm UV}\approx17$, $s_{\rm opt}\approx0$ & $f_{\rm eff}\approx0.04$ field & UV traces $<300$ Myr, per-event $-0.22$ & field negligible, cluster-only $\Delta\log R_0^{\rm UV}=-0.09$ \\ 
\midrule
\multicolumn{5}{@{}l}{\textit{Quiescent galaxies}}\\ \midrule
Dry major mergers & $s=1$ & $\sim0.8$ events, $\mu(M_0)=0.18$ & UV$\approx$opt (added stars old) & $\Delta\alpha=+0.03$ ($+0.02$--$0.08$), $\Delta\log R_0=+0.08$ (modest) \\ \midrule
Minor mergers, accretion & $s=2.3$ & $\mu(M_0)=0.014$ & track steeper in UV $s_{\rm UV}\approx3.7$ & $\Delta\alpha=+0.05$, $\Delta\log R_0=+0.03$, UV$>$opt $+0.02$ (significant) \\ \midrule
Adiabatic expansion & $R_f/R_i=1/(1-f)$, $s\approx-\eta(1-f_{\rm DM})$ & stellar return $f\approx0.11$, $\eta=0.5$ & UV$\approx$opt (homologous) & $\Delta\alpha=-0.006$, $\Delta\log R_0=+0.02$ (negligible) \\ \midrule
High-velocity encounters & $s=0.7$ ($0.4$--$1.1$) & $f_{\rm aff}\lesssim0.008$ & UV$\approx$opt (uniformly old) & $\Delta\alpha,\Delta\log R_0\approx0$, $s \approx \alpha$ (negligible) \\ \midrule
Tidal stripping & $s_{\rm opt}\approx1.7$, $s_{\rm UV}\approx2.0$ (truncating an $n=4$ profile) & $\langle\Delta\log\mstar\rangle=-0.07$, $f_{\rm aff}=f_{\rm sat}\times f_{\rm strip}$ & metal-poor UV envelope removed & $\Delta\alpha=+0.01$, $\Delta\log R_0=-0.005$, UV$>$opt (negligible) \\ \midrule
Progenitor bias & mixture $f_{\rm new}(M) \Delta(M)$, not a track & $f_{\rm new}=0.67/0.44/0.23$ & UV fades $<1$ Gyr, adopted $w \approx 0.7$ & $\Delta\alpha=-0.12$ (flatten), $\Delta\log R_0=+0.02$ at $M_0$, UV$<$opt (significant) \\
\bottomrule
\end{tabularx}
\end{adjustwidth}
\end{table}

\textit{High-velocity encounters.} From Equation~(\ref{eq:impulse}) the susceptibility scales as $r_h^{3}/m$, which favours low-mass extended disks. We therefore taper the effect above $\log\mstar/\msun=10$ with the factor $\tfrac{1}{2}[1-\tanh((\log\mstar-10)/0.15)]$, following the decline of stripping efficiency toward higher mass reported by \citet{boselliEnvironmentalEffectsLateType2006} and \citet{boselliOriginDwarfEllipticals2008}. Summing the tidal-shock energy from each encounter along a realistic cluster orbit gives a cumulative $\Delta E/|E|$ well below unity for a median orbit, consistent with the $N$-body result that more than three quarters of orbits produce no stellar mass loss even for the loosely bound early-type dwarfs of Virgo-like clusters~\citep{smithSensitivityHarassmentOrbit2015}. For~QGs the removal scale is the King tidal radius $r_t\simeq R_{\rm peri}[m/3M(<R_{\rm peri})]^{1/3}$~\citep{kingStructureStarClusters1962}, where $R_{\rm peri}$ is the pericentric distance and $M(<R_{\rm peri})$ the host mass enclosed within it. Massive QGs above the pivot sit in deeper potentials than those dwarfs; hence, $r_t$ remains well outside their stellar body on a typical orbit and a single encounter strips at most a few per cent of their stars. The~QG track slope $s\approx0.7$ nearly matches the relation slope above the pivot, which pushes the imprint toward zero independently of the population~dilution.

\textit{Minor mergers and accretion.} For dry spheroids the accreted low-density satellites are deposited at a large radius and grow in size as $\re\propto\mstar^2$ ($s\approx2$ in the energy argument for accreted material with negligible internal binding; \citep{naabMinorMergersSize2009}), and~$N$-body experiments of 1:5 mass ratio mergers give $s\approx2.3$ once the satellites retain their dark-matter haloes \citep{hilzHowMinorMergers2013}. We adopt $s=2.3$ for QGs. This dry-spheroid scaling does not carry over to gas-rich disk hosts, where accreted gas can sink to the centre and ongoing in situ star formation dilutes the size growth. For~SFGs we therefore adopt a lower, low-confidence $s\approx1$, bracketed by the dry-spheroid limit ($s=2$) and a strong central-star-formation offset ($s=0.5$; \citep{kavirajImportanceMinormergerdrivenStar2014}).

\textit{Tidal stripping of \qgs.} We strip a galaxy by truncating an $n=4$ S\'{e}rsic profile at radius $R_{\rm cut}=x \re$ and solving $L(<R_{\rm new})=\tfrac{1}{2}L(<R_{\rm cut})$ from the curve of growth $L(<x)\propto\gamma(2n,b_n x^{1/n})$, where $\gamma$ is the lower incomplete gamma function and $b_n$ the S\'{e}rsic constant~\citep{grahamConciseReferenceProjected2005}. Averaged over retained-mass fractions of $0.70$--$0.95$ this gives a track slope $s\approx1.7$ for $n=4$, and~$\approx1.4$ for $n=2.5$. The~UV light is more extended ($R_{\rm UV}\approx1.15 R_{\rm opt}$) and hence, the~same physical cut removes a larger UV fraction and the per-galaxy contrast is ${\rm d}\log\re^{\rm UV}/{\rm d}\log\re^{\rm opt}\approx1.2$ \citep{pulsoniStellarHalosETGs2021}. The~affected fraction $f_{\rm sat}f_{\rm strip}$, the~product of the satellite fraction and the stripped-satellite fraction, dilutes this to $\Delta\log R_0\approx-0.005$ dex at the population~level.

\textit{Adiabatic expansion of \qgs.} Equation~(\ref{eq:adiab}) gives expansion factors $R_f/R_i=1.4$--$2.0$ for a removed fraction $f=0.3$--$0.5$, the~range set by the gas-rich quenching event~\mbox{\citep{fanDramaticSizeEvolution2008,ragone-figueroaPuffingEarlytypeGalaxies2011}}. The~impulsive form unbinds the system at $f=0.5$. For~already-quiescent galaxies at $z<1$ the only steady expellable reservoir is stellar-evolution mass return, $f\approx0.11$ at the fiducial mass~\citep{leitnerFuelEfficientGalaxies2011}, driving a near-vertical effective track $s\approx-\eta (1-f_{\rm DM})$ with expulsion efficiency $\eta\lesssim1$ and dark-matter fraction $f_{\rm DM}\approx0.15$. This returns $\Delta\log R_0\approx+0.02$ dex over $z=1 \to 0$, at~or below the measurement~yardstick.

\textit{Progenitor bias.} This is the one channel modelled as a change in the population mixture rather than a displacement of individual galaxies. At~each mass a fraction \mbox{$f_{\rm new}(M)=1-1/g(M)$} of today's QGs quenched since $z=1$, where $g(M)=n_Q(M,0)/n_Q(M,1)$ is the quiescent number-density growth. We adopt $g\approx3/1.8/1.3$ at $\log\mstar/\msun=10.0/10.7/11.3$~\citep{moustakasPRIMUSConstraintsStar2013,moutardVIPERSMultiLambdaSurvey2016}, which gives $f_{\rm new}=0.67/0.44/0.23$. These newcomers enter at their quench-epoch star-forming size, larger at fixed mass than the incumbent \qgs. The~relation therefore shifts by $\Delta\log R_0(M)=f_{\rm new}(M) \Delta(M)$, with~$\Delta(M)$ the mean newcomer size offset. Because~$f_{\rm new}$ falls with mass, the~low-mass end is lifted most and the slope flattens. Each newcomer stays UV-bright only while its disk survives, for~under $1$~Gyr after quenching, and~fades to a mild excess afterward. Averaged over arrival times this gives a mean UV weight $w\approx0.7$ and a weak UV-minus-optical signal below $0.01$~dex whose sign is not robust. The~result is a slope change $\Delta\alpha=-0.12$ with a zero-point lift $\Delta\log R_0=+0.02$~dex at the fiducial~mass.


\begin{adjustwidth}{-\extralength}{0cm}
\reftitle{References}

\PublishersNote{}
\end{adjustwidth}
\end{document}